\documentclass{aa} 
\usepackage{graphicx,natbib,psfig,textcomp,fontenc,amssymb}  
\newcommand{\ltsima} {$\; \buildrel < \over \sim \;$} 
\newcommand{\gtsima} {$\; \buildrel > \over \sim \;$} 
\newcommand{\lta} {\lower.5ex\hbox{\ltsima}} 
\newcommand{\gta} {\lower.5ex\hbox{\gtsima}} 
\newcommand{\Ha} {H$\alpha$}

\newcommand{\ergs}{\>{\rm erg}\,{\rm s}^{-1}}
\newcommand{\ergsHz}{\ensuremath{{\rm erg}\,{\rm s}^{-1}\,{\rm Hz}^{-1}}}

\newcommand{\oiii} {[O~III]}  
\newcommand{\tc} {$\theta_{\rm c}$}

\begin{document}

\title{An optical spectroscopic survey of the 3CR sample of radio galaxies
  with $z<0.3$. \\ V. Implications for the unified model for FR~IIs.}
  
\titlerunning{Implications for the unified model for FR~IIs.} 
\authorrunning{Baldi, R.D. et al.}
  
\author{Ranieri D. Baldi \inst{1} \and Alessandro Capetti \inst{2} \and Sara
  Buttiglione \inst{3} \and Marco Chiaberge \inst{4,5,6} \and Annalisa Celotti
  \inst{1,7,8} }
   

\institute{SISSA-ISAS, via Bonomea 265, I-34136 Trieste, Italy \and INAF -
  Osservatorio Astrofisico di Torino, Strada Osservatorio 20, I-10025 Pino
  Torinese, Italy, \and INAF - Osservatorio Astronomico di Padova, Vicolo
  dell'Osservatorio 5, I-35122 Padova, Italy, \and Space Telescope Science
  Institute, 3700 San Martin Drive, Baltimore, MD 21218, USA, \and
  INAF-Istituto di Radio Astronomia, via P. Gobetti 101, I-40129 Bologna,
  Italy, \and Center for Astrophysical Sciences, Johns Hopkins University,
  3400 N. Charles Street Baltimore, MD 21218, \and INAF-Osservatorio
  Astronomico di Brera, via E. Bianchi 46, 23807 Merate, Italy, \and INFN-
  Sezione di Trieste, via Valerio 2, 34127, Trieste, Italy}

\date{}

\abstract{We explore the implications of our optical spectroscopic survey of
  3CR radio sources with z $<$ 0.3 for the unified model (UM) for radio-loud
  AGN, focusing on objects with a 'edge-brightened' (FR~II) radio
  morphology. The sample contains 33 high ionization galaxies (HIGs) and 18
  broad line objects (BLOs). According to the UM, HIGs,
  the narrow line sources, are the nuclearly obscured counterparts of BLOs. \\
  The fraction of HIGs indicates a covering factor of the circumnuclear matter
  of 65\% that corresponds, adopting a torus geometry, to an opening angle of
  $50^{\circ} \pm 5$. No dependence on redshift and luminosity on the torus
  opening angle emerges. We also consider the implications for a 'clumpy' torus.\\
  The distributions of total radio luminosity of HIGs and BLOs are not
  statistically distinguishable, as expected from the UM. Conversely, BLOs
  have a radio core dominance, $R$, more than ten times larger with respect to
  HIGs, as expected in case of Doppler boosting when the jets in BLOs are
  preferentially oriented closer to the line of sight than in HIGs. Modeling
  the $R$ distributions leads to an estimate of the
  jet bulk Lorentz factor of $\Gamma \sim 3 - 5$. \\
  The test of the UM based on the radio source size is not conclusive due to the
  limited number of objects and because the size distribution is dominated by
  the intrinsic scatter rather
  than by projection effects.  \\
  The [O~II] line luminosities in HIGs and BLOs are similar but the [O~III]
  and [O~I] lines are higher in BLOs by a factor of $\sim$ 2. We ascribe
  this effect to the presence of a line emitting region located within the
  walls of the obscuring torus, visible in BLOs but obscured in HIGs, with a
  density higher than the [O~II] critical density. We find evidence that BLOs
  have broader [O~I] and [O~III] lines than HIGs of similar [O~II] width, as
  expected in the presence of high density gas in the proximity of the central
  black hole. \\
  In conclusion, the radio and narrow line region (NLR) properties of HIGs and
  BLOs are consistent with the UM predictions when the partial obscuration of
  the NLR is
  taken into account. \\
  We also explored the radio properties of 21 3CR low ionization galaxies
  (LIGs) with a FR~II radio morphology at z$<$0.3. We find evidence that they
  cannot be part of the model that unifies HIGs and BLOs, but they are instead
  intrinsically different source, still reproduced by a randomly oriented
  population.

\keywords{galaxies: active, galaxies: jets, galaxies: nuclei} }

\maketitle
  
\section{Introduction}
\label{intro}

The unified model (UM) for active galactic nuclei (AGN) postulates that
different classes of objects might actually be intrinsically identical and
differ solely for their orientation with respect to our line of sight (see,
e.g., \citealt{antonucci93} for a review). The origin of the aspect dependent
classification is due to the presence of: i) circumnuclear absorbing material
(usually referred to as the obscuring torus) that produces selective
absorption when the source is observed at a large angle from its radio axis;
ii) Doppler boosting associated with relativistic motions in AGN jets.

According to this model, for radio-loud AGN a source appears as a quasar only
when its radio axis is oriented within a small cone opening around the
observers line of sight \citep{barthel89}. In the unification scheme of radio
loud AGN (e.g., \citealt{urry95}) narrow-lined radio galaxies of FR~II type
\citep{fanaroff74} and broad-lined FR~IIs together with RL QSOs, also called
Broad Line Objects (BLO), are considered to be intrinsically
indistinguishable.  Their different aspect (in particular the absence of broad
emission lines in FR~IIs) is only related to their orientation in the sky with
respect to our line of sight. Therefore, the UM, in its stricter
interpretation of a pure orientation scheme, predicts that narrow-line and
broad-line FR~II are drawn from the same parent population.  Among the several
pieces of evidence in favor of the UM, probably the most convincing one is the
detection of broad lines in the polarized spectra of narrow-line objects
\citep{antonucci82,antonucci84} interpreted as the result of scattered light
from an otherwise obscured nucleus.

The FR~II radio galaxies population consists of two main families, based on
the optical narrow emission-line ratios: high ionization galaxies (HIG) and
low ionization galaxies (LIG)\footnote{We adopt the HIG/LIG nomenclature that
  better represents the separation of the two classes based on the ionization
  conditions in the narrow-line region gas in these objects. This
  classification is however entirely consistent with the HEG/LEG scheme
  (high/low excitation galaxies) widely adopted in the literature.}
\citep{hine79,jackson97,buttiglione10}. Such a dichotomy also corresponds to a
separation in nuclear properties at different wavelengths (e.g.,
\citealt{chiaberge02,hardcastle06,baldi10b}).

In \citet{buttiglione10} we noted that all BLOs have a high ionization
spectrum, based on the ratios between narrow emission lines. Since this
spectroscopic classification should not depend on orientation (an issue that,
however, we will discuss later in more detail) this suggests to consider the
narrow-lined HIGs as the nuclearly obscured counterpart of the BLO population.
On the other hand, LIGs appear to be a different class of AGN. In fact
\citet{laing94} have pointed out that LIGs are unlikely to appear as quasars
when seen face-on and that these should be excluded from a sample while
testing the unified scheme model (see also \citealt{wall97} and
\citealt{jackson99}). Therefore, in order to test the validity of the UM for
RL AGN, it is necessary to treat separately FR~II HIGs (with BLOs) and LIGs
because of their different nuclear properties and spectroscopic
classifications.

The Revised Third Cambridge Catalog 3CR
\citep{bennett62b,bennett62a,spinrad85} is perfectly suited to test the
validity and to explore the implications for the UM for local powerful
radio-loud AGN. This is because it is based on the low frequency radio
emission which should make it free from orientation biases. Furthermore, the
results of a complete optical spectroscopic survey obtained for the 3CR radio
sources, limited to $z<0.3$ \citep{buttiglione09,buttiglione10,buttiglione10b}
gives a robust spectral classification of all objects.

The completeness and the homogeneity of the sample, reached with the 3CR
  spectroscopic survey, are fundamental for obtaining results with a high
  statistical foundation. Thanks to these conditions, our intention is to test
  various predictions and to discuss the implications of the UM on the
  properties of the sample of 3CR FR~II radio galaxies with results more
  robust than in previous works.

The UM validity, that can be tested with the available data, is
  represented by the consistency of the isotropic properties of the two
  sub-samples of HIGs and BLOs. According to the 'zeroth-order approximation'
  of the AGN UM, the extended radio emission and the NLR are supposed to be
  insensitive to orientation. The extended radio emission is the main
characteristic of radio-loud AGN: it is isotropic, it extends far beyond the
host galaxy dimensions, and it is not affected by absorption. The narrow lines
are observed in all AGNs (except in BL~Lac objects, where the beamed nuclear
emission dominates the spectrum, diluting their intrinsically weak emission
lines, \citealt{capetti10}) and their extent, up to kpc scales, is certainly
larger than the size of the torus and they cannot be completely obscured. Thus
these quantities can be compared in BLOs and HIGs to test the UM. Another
important aspect of orientation is an apparent change of the radio core
luminosity: since the core emission is subject to relativistic beaming
effects, when the jet is pointing in a direction close to our line of sight
(face-on AGN) its emission will be enhanced while, in the case the AGN is
edge-on the core emission, should be de-beamed. The separation in radio core
power between HIGs and BLOs can be modeled to extract information on the jet
properties. In addition, the orientation scheme has also the geometric effect
of producing the projected sizes of BLOs in radio images smaller than those of
HIGs. Such a difference can be also used to test the validity of the UM for RL
AGN. 

Finally, we consider the FR~II LIG sub-sample to study how they fit into in
the oriented-based unified scheme, bearing in mind that \citet{laing94}
already suggest that LIGs appear to be a intrinsically different RL AGN
population from HIGs and BLOs.

The structure of this paper is as follows: in Sect. \ref{thesample} we define
the sample considered and list the main properties of the sources; the
separation between BLOs and HIGs is critically reviewed in
Sect. \ref{step1}. In Sect. \ref{step3} we derive the geometric properties of
the obscuring material. The implications of the UM on the radio and emission
line properties are presented in Sect. \ref{step5} through \ref{step6}. The
role of LIGs in the UM is addressed in Sect. \ref{lstep}. The results are
discussed in Sect. \ref{discap5} and summarized in Sect. \ref{concl5}, where
we also draw our conclusions.

\section{The sample}
\label{thesample}

We consider all the 3CR FR~II radio sources with a limiting redshift of
$z<0.3$ and an optical spectrum characterized by emission lines of high
ionization. As explained in the Introduction, HIGs and BLOs differ by
  definition for the presence of broad emission lines in their optical
  spectra. Finally, we also consider separately the 3CR FR~II LIG sources,
again with $z<0.3$.

The main data for these sources are reported in Table \ref{sample}.  The
spectroscopic data are taken from \citet{buttiglione09,buttiglione10b}, while
the spectral classification is from \citet{buttiglione10,buttiglione10b}. The total radio
luminosities at 178 MHz are from \citet{spinrad85}, while the radio core power
at 5 GHz are from the compilation of literature data presented by \citet{baldi10b}.

\begin{table*}
  \begin{center}
    \caption{Main properties of the sample of HEGs and BLOs with z$<$0.3 in
      the 3CR catalog.}
  \label{sample}
  \begin{tabular}{l|c | c c |c c c | c}
    \hline \hline
 Name &  z &  Log P$_{core}$ &  Log $L_{\rm 178}$  & L$_{\rm [O~III]}$ & Log $L_{\rm [OII]} $  &   Log $L_{\rm [OI]}$  &  class \\
    \hline
3C~017   & 0.220 & 32.94 &  34.45 &  41.99 &  41.89  &  41.53   & BLO \\
3C~018   & 0.188 & 32.00 &  34.27 &  42.55 &  41.96  &  41.56   & BLO \\
3C~020   & 0.174 & 30.44 &  34.55 &  41.54 &  41.52  &  40.55   & HEG \\
3C~033   & 0.059 & 30.36 &  33.65 &  42.18 &  41.88  &  41.03   & HEG \\
3C~033.1 & 0.180 & 31.19 &  34.07 &  42.29 &  41.64  &  41.24   & BLO \\
3C~061.1 & 0.184 & 30.49 &  34.47 &  42.47 &  42.01  &  40.95   & HEG \\
3C~063   & 0.175 & 31.12 &  34.21 &  41.63 &  41.51  &  40.89   & HEG \\
3C~079   & 0.255 & 31.39 &  34.78 &  42.86 &  42.10  &  41.08   & HEG \\
3C~093.1 & 0.243 &       &  34.24 &  42.67 &  42.68  &  41.81   & HEG \\
3C~098   & 0.030 & 29.87 &  32.99 &  41.00 &  40.94  &  39.71   & HEG \\
3C~105   & 0.089 & 30.46 &  33.54 &  41.45 &  40.83  &  40.48   & HEG \\
3C~111   & 0.048 & 31.77 &  33.54 &  42.44 &  41.74  &  41.03   & BLO \\
3C~133   & 0.277 & 32.53 &  34.72 &  42.76 &  42.67  &  41.79   & HEG \\
3C~135   & 0.125 & 30.31 &  33.84 &  42.05 &  41.61  &  40.77   & HEG \\
3C~136.1 & 0.064 & 29.16 &  33.13 &  41.45 &  40.98  &  40.07   & HEG \\ 
3C~171   & 0.238 & 30.55 &  34.51 &  42.88 &  42.72  &  41.82   & HEG \\
3C~180   & 0.220 &       &  34.32 &  42.34 &  41.71  &  40.78   & HEG \\
3C~184.1 & 0.118 & 30.37 &  33.66 &  42.22 &  41.50  &  40.70   & BLO \\
3C~192   & 0.059 & 29.82 &  33.25 &  41.35 &  41.08  &  39.99   & HEG \\
3C~197.1 & 0.127 & 30.43 &  33.55 &  40.94 &  40.64  &  40.21   & BLO \\
3C~219   & 0.174 & 31.69 &  34.53 &  41.77 &  41.43  &  41.19   & BLO \\
3C~223   & 0.136 & 30.70 &  33.85 &  42.17 &  40.66  &  40.95   & HEG \\
3C~223.1 & 0.107 & 30.36 &  33.23 &  41.57 &  40.73  &  39.96   & HEG \\
3C~227   & 0.085 & 30.58 &  33.74 &  41.76 &  40.82  &  40.43   & BLO \\
3C~234   & 0.184 & 32.04 &  34.47 &  43.11 &  42.10  &  41.28   & HEG \\
3C~273   & 0.158 & 34.34 &  34.62 &        &         &          & BLO \\
3C~277.3 & 0.085 & 30.34 &  33.21 &  40.94 &  40.90  &  40.29   & HEG \\
3C~284   & 0.239 & 30.44 &  34.28 &  41.59 &  40.93  &  40.02   & HEG \\
3C~285   & 0.079 & 30.03 &  33.23 &  40.55 &  40.75  &  39.68   & HEG \\
3C~287.1 & 0.215 & 32.71 &  34.04 &  41.74 &  41.63  &  41.18   & BLO \\
3C~300   & 0.272 & 31.27 &  34.60 &  42.00 &  41.95  &  40.95   & HEG \\
3C~303   & 0.141 & 31.94 &  33.77 &  41.74 &  41.53  &  40.94   & BLO \\
3C~303.1 & 0.269 & 31.04 &  34.25 &  42.41 &  42.27  &  41.48   & HEG \\
3C~305   & 0.041 & 30.07 &  32.79 &  41.03 &  40.84  &  40.16   & HEG \\
3C~321   & 0.097 & 30.89 &  33.49 &  40.90 &  39.84  &  39.30   & HEG \\ 
3C~323.1 & 0.264 & 31.89 &  34.31 &  42.80 &  41.96  &  41.33   & BLO \\
3C~327   & 0.104 & 30.99 &  33.98 &  42.24 &  41.43  &  40.87   & HEG \\
3C~332   & 0.151 & 30.79 &  33.77 &  41.81 &  41.29  &  40.64   & BLO \\
3C~379.1 & 0.256 & 30.90 &  34.16 &  41.85 &  41.15  &  40.73   & HEG \\
3C~381   & 0.160 & 30.63 &  34.06 &  42.37 &  41.75  &  40.82   & HEG \\
3C~382   & 0.057 & 31.22 &  33.19 &  41.77 &  40.92  &  40.68   & BLO \\ 
3C~390.3 & 0.056 & 31.46 &  33.54 &  42.08 &  40.59  &  41.01   & BLO \\
3C~403   & 0.059 & 29.96 &  33.16 &  41.75 &  40.81  &  40.32   & HEG \\
3C~410   & 0.248 & 33.44 &  34.80 &  42.03 &  40.88  &  41.12   & BLO \\
3C~433   & 0.101 & 30.11 &  34.16 &  41.67 &  41.25  &  40.73   & HEG \\
3C~436   & 0.214 & 31.39 &  34.37 &  41.56 &  41.28  &  34.12   & HEG \\
3C~445   & 0.056 & 31.42 &  33.26 &  42.50 &  41.47  &  41.07   & BLO \\
3C~452   & 0.081 & 31.34 &  33.94 &  41.34 &  41.12  &  40.59   & HEG \\
3C~456   & 0.233 & 31.57 &  34.23 &  42.81 &  42.45  &  41.66   & HEG \\
3C~458   & 0.289 &       &  34.58 &  42.03 &         &          & HEG \\
3C~459   & 0.220 & 33.20 &  34.55 &  42.04 &  42.29  &  41.26   & BLO \\
   \hline
 \end{tabular}\\
 \end{center}
Line luminosities are in units of erg s$^{-1}$, while radio luminosities are
in erg s$^{-1}$ Hz$^{-1}$. 
\end{table*}

\section{Are the broad lines really missing in the HIGs spectra?}
\label{step1}

\begin{figure*}
  \centerline{ 
    \psfig{figure=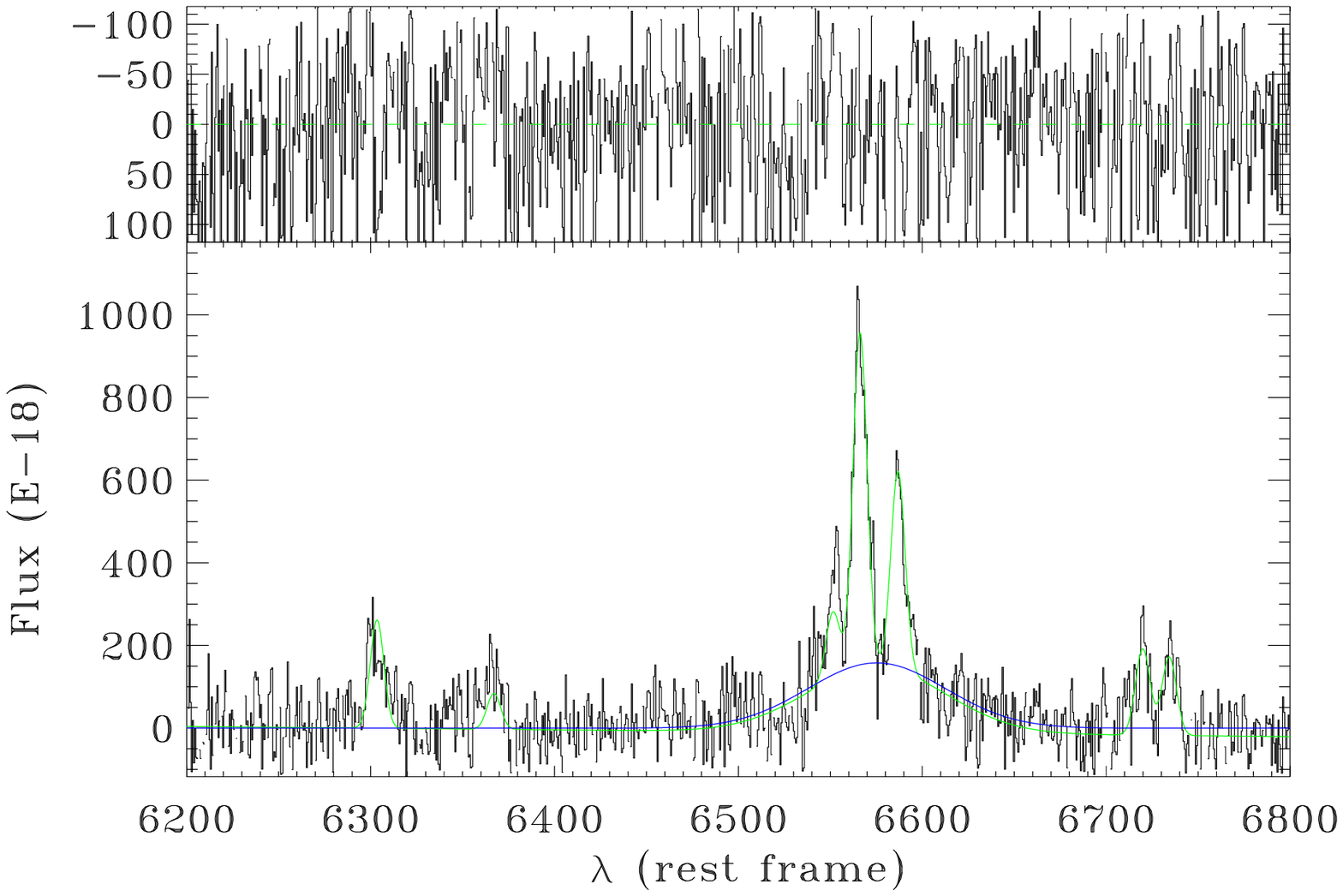,width=0.5\linewidth}
    \psfig{figure=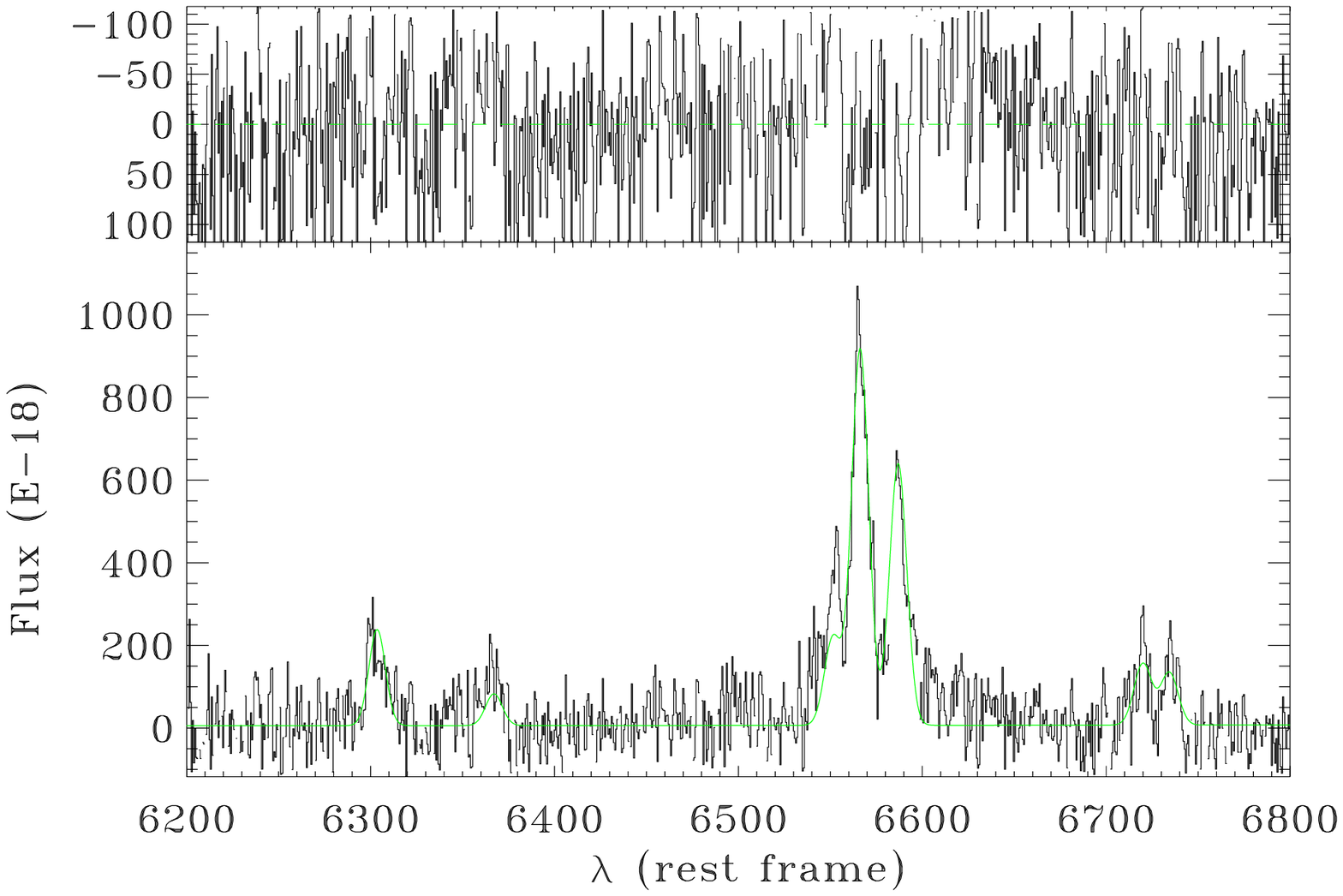,width=0.5\linewidth}
 }
 \caption{\label{3c133blr} In the left panel we show the fit to the spectrum
   of 3C~133 obtained by forcing the presence of a broad \Ha, while in the
   right panel we show the result of the fit with only narrow line
   components. The original spectrum is in black, the narrow lines are in
   green, and the broad \Ha\ is in blue. The residuals are shown in the top
   panel of both figures.}
\end{figure*}

BLOs are defined as the objects in which emission from a BLR is clearly
detected in the optical spectra in the form of a broad Balmer line emission
underlying the forbidden narrow emission lines; a fit including only narrow
components leaves strong positive residuals. Conversely, a BLR is apparently
missing in HIGs. But is it possible that lines are not detected in HIGs broad
only because of observational limitations, such as they are hidden by the
continuum emission or by the narrow emission lines, and/or they are not
visible just due an insufficient quality of the spectra? In other words, how
reliable is the separation between BLOs and HIGs?

In order to answer this question we looked more closely for broad lines
footprints in HIGs. We focused on the \Ha\ emission line because it is the
brightest permitted emission lines in the optical spectra. We used the {\it
  specfit} package from the IRAF data reduction software forcing the program
to fit a broad line underlying the complex of [N~II]+\Ha\ emission lines even
in absence of clear residuals. We tested different line widths, fixing its
value in {\it specfit} within the range 4000-8000 km s$^{-1}$ as measured in BLOs.
We then assessed the reliability of the presence of the broad \Ha\ line with a
likelihood ratio test (F-test, \citealt{bevington03}). This compares the
residuals of the fit with and without an additional component in the model,
taking into account the increased number of degrees of freedom. By setting a
significance threshold at 95\% a broad line is not detected in any
HIGs with the exception of 3C~234. We then set as upper limit to the broad
\Ha\ flux 3 times the measurement errors. As an example we report in
Fig. \ref{3c133blr} the case of 3C~133, the object with the largest allowed
broad line flux.

\begin{figure}[ht]
  \centerline{ 
    \psfig{figure=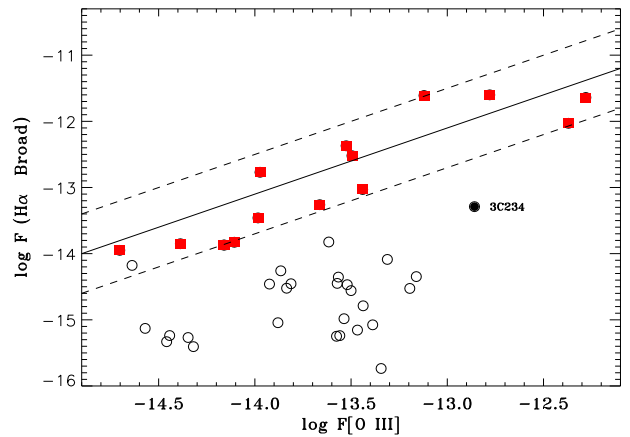,width=1.0\linewidth}
 }
 \caption{\label{bloheg2} Comparison between the narrow \oiii\ line
   and the broad \Ha\ fluxes. Red squares are BLOs, black circles
   are HIGs, with empty symbols representing upper limits. The fluxes are in
   erg cm$^{-2}$ s$^{-1}$. The solid line marks the average ratio between the
   two emission lines as measured in BLOs, ($F_{\rm H\alpha \,\ broad}/F_{\rm
     [O~III]} \sim 8$). The dashed lines illustrates a change of a factor of 4 in
   this value. }
\end{figure}

The result of this analysis is shown in
Fig.~\ref{bloheg2}. \citet{buttiglione10} noted that for BLOs there is a
proportionality between the broad and narrow emission line fluxes, as a
consequence of the fact that both originate from the same ionizing
radiation. The average relation between the \oiii\ and \Ha\ broad line fluxes
derived from the BLOs is reported as the solid line in the
Fig.~\ref{bloheg2}. Conversely, HIGs are all (but one) located well below (by
a factor of 10-1000) the relation defined by BLOs. The only object consistent
with the broad \Ha\ - \oiii\ ratio of BLOs is 3C~284, whose spectrum is of
rather poor quality and has the lowest value of \oiii\ flux.  We conclude that
the broad \Ha\ line flux in HIGs is significantly lower than it would have
been expected based on their \oiii\ flux. This implies that any broad line in
HIGs does not obey to the same scaling of BLOs and that the separation between
the two classes is robust and reliable.

The only exception to this scheme is 3C~234 where a BLR is seen, although
$\sim$ 20 times fainter that one would predict from its \oiii\ flux.  However,
spectro-polarimetric studies of this source revealed that its broad lines are
highly polarized and they are the ascribed to of scattered nuclear light
\citep{antonucci84}. This is actually one of the observational results at the
very foundation of the UM.  In 3C~234 we do not have a direct view of the BLR,
but given its high flux, the small scattered fraction is sufficient to see a
broad line in its total intensity spectrum.

\section{Geometry of the obscuring torus}
\label{step3}

Considering that our sample is complete for redshift $z<0.3$ and not subject
to selection biases, we can use the ratio between the number of BLOs and HIGs
to study the geometry of the  circumnuclear obscuring material. This analysis
implicitly assumes the validity of the UM, i.e., that the differences between
HIGs and BLOs are solely due to a different orientation with respect to the
line of sight and to the presence of selective obscuration. We initially limit
ourselves to a simple structure, assuming that it produces complete
obscuration toward the nucleus when its axis forms with the line of sight an
angle larger than a critical value (\tc), while it leaves a free view for
smaller angles, i.e., that it takes the form of a torus with sharp boundaries.

For angles smaller than \tc\ we then expect to look inside the torus, to
see the BLR and thence to observe a BLO; for angles larger than \tc\ the BLR is
obscured by the torus, thus we observe a HIG.  For a randomly oriented set of
sources, the probability of finding an object within the cone with an opening
angle \tc\ is $P(\theta <$ \tc)=1 - cos \tc \citep{barthel89}. In the
complete 3CR sample for $z<0.3$ we have 18 BLOs and 33 HIGs from the optical
spectroscopic classification.  The resulting value for the critical angle is
\tc = 49.7$^\circ$.

In order to estimate the uncertainty on \tc\ related to the limited size of
our sample, we ran a set of Monte Carlo simulations. More specifically, we
measured the distribution of the number ratio between HIGs and BLOs in 1000
realizations of a sample of 51 randomly oriented sources and derived the
corresponding value of \tc. This procedure yields a dispersion of 5$^\circ$.
The final result is then \tc = $50^{\circ} \pm 5^{\circ}$.

We also split the radio sources in two sub-samples depending on the redshift.
Since we are considering a flux limited sample this generally corresponds into
splitting the sample at different levels of luminosity. This is aimed at
verifying whether \tc\ is constant with luminosity or, conversely, it changes
for sources at higher redshift/luminosity. We have chosen a threshold that
divides almost equally the objects of the sample, i.e., $z = 0.15$. The median
luminosities of the sources in the two redshift bins differ by almost an order
of magnitude, being log $L_{\rm 178 \, MHz} = 33.5$ $\ergsHz$ and log $L_{\rm
  178 \, MHz} = 34.4$ $\ergsHz$ for $z<0.15$ and $z>0.15$, respectively. In
Table \ref{hegblo} we report the numbers of HIGs and BLOs in the two redshift
bins. We repeated the Monte Carlo simulations described above and we found
that the critical angle estimates for the two subclasses are \tc$ = 48^{\circ}
\pm 7^{\circ}$ for galaxies with $z \leq 0.15$ and \tc$ = 51^{\circ} \pm
7^{\circ}$ for $0.15<z<0.3$.  These values are consistent with each other
within the errors and also with the estimate for the whole sample.  Thus, we
find no evidence for a change in the torus structure with luminosity and
redshift.

\begin{table}
  \begin{center}
    \caption{HIGs and BLOs in redshift subclasses}
  \label{hegblo}
  \begin{tabular}{c| c c c| c}
    \hline \hline
     $z$ interval &HIGs & BLOs & Total & BLOs/total\\
    \hline
$z \leq 0.15$ &16  & 8  & 24  & 33\% \\
$z   >  0.15$ &17  & 10  &27  & 37\% \\
    \hline                  
    Total     &33  &18  &51  & 35\% \\
    \hline
  \end{tabular}\\
  \end{center}
\end{table}

\begin{figure}
  \centerline{ \psfig{figure=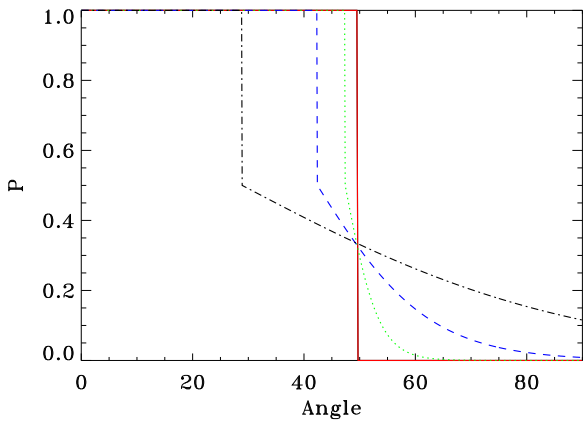,width=0.99\linewidth}
 }
 \caption{\label{torush3} Probability distribution adopted to simulate the
   effect of a clumpy torus. $P$ is the probability of an AGN seen at angle
   $\theta$ (in degrees) to be classified as a BLO. The various curves
   corresponds to $t = 0.01^\circ$ (solid red), $3^\circ$ (dotted green),
   $10^\circ$ (dashed blue), and $30^\circ$(dot dashed black), see
   text. $\theta_{c}$ is the 'knee' angle where the probability drastically
   changes slope.}
\end{figure}

We also considered the possibility of a ``clumpy'' torus. This consists in a
toroidal structure around the AGN made of clumps that block the nuclear
radiation according to a probabilistic law. In such a case, there is a non
zero probability to observe a BLO even for viewing angles $\theta > $ \tc\
(see Fig.~\ref{torush3}). We adopted the probabilistic function

$$P(\theta,t) = 1/(1+e^{\frac{(\theta - \theta_{c})}{t}}) \,\,\,\, \rm{for}
\,\,\,\,\, \theta>\theta_{c}$$
 
whose limit for small values of the parameter $t$ is the Heaviside step
function used before. \tc\ is varied for each value of $t$ to reproduce the
observed fractions of HIGs and BLOs. For example, for $t=10^\circ$, we derive
\tc $\sim 43^\circ$ and there is a $\sim$0.9\% probability to have a clear
view of the nucleus for an object seen along the torus equator, $P_{90}$. The
results for $t=3^\circ$ are \tc $\sim 47^\circ$ and $P_{90}
\sim7\times10^{-7}$.

The values of $P_{90}$ obtained for $t$ = 10$^{\circ}$ and 3$^{\circ}$,
corresponds to a number of clumps along the line of sight of $\cal N \sim $ 5
- 15, respectively \citep{natta84}, assuming the torus as a thick inhomogeneous
dust layer in front of an extended emitting source.  According to the analysis
of \citet{nenkova08b}, this is the range in the number of dusty clouds along
radial equatorial rays that accounts for the AGN infrared observations. From
this simple approach, we conclude that the torus critical angle changes only
slightly in case the torus has a clumpy structure. We will show that this has
no significant effects on the results obtained in the following sections.

\section{Radio properties and the unified model}
\label{step5}

\begin{figure*}
  \centerline{ \psfig{figure=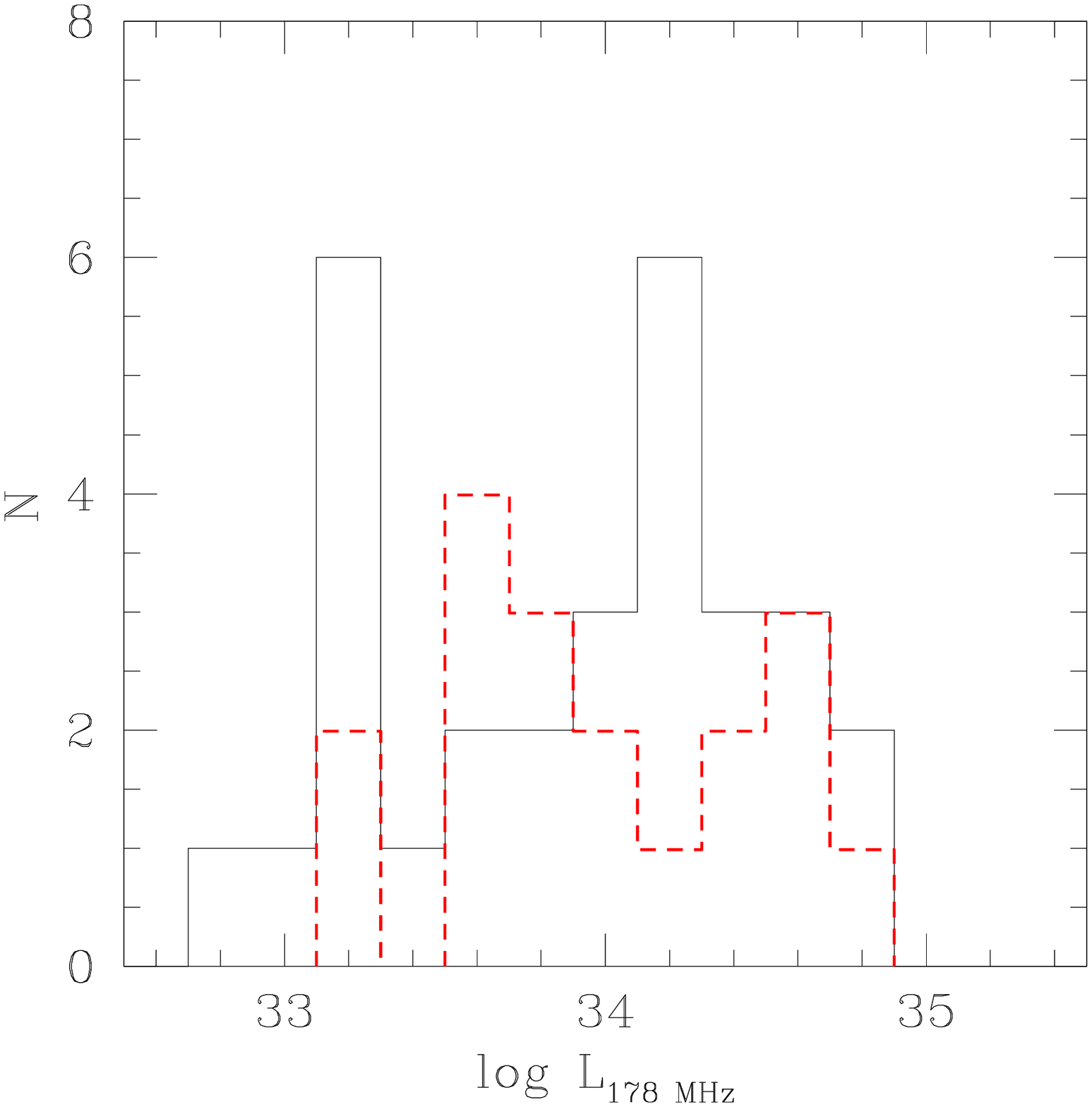,width=0.5\linewidth}
   \psfig{figure=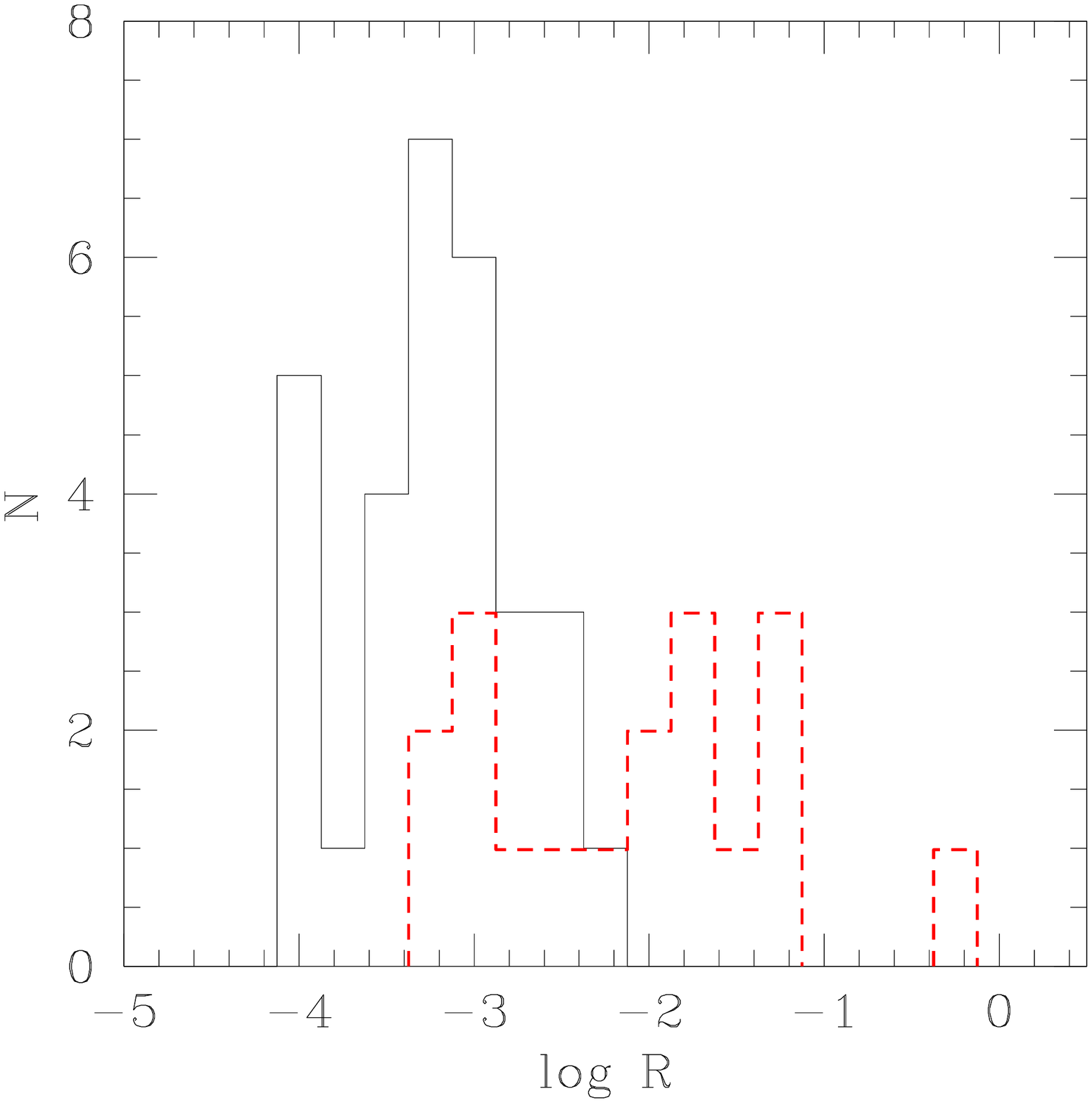,width=0.50\linewidth}}
 \caption{\label{isto} Histograms comparing the distribution of the 178 MHz
   radio luminosity in erg s$^{-1}$ Hz$^{-1}$ units (left) and core dominance
   $R$ (right) for HIGs (solid black) and BLOs (dashed red). $R$ is the ratio
   between the total radio luminosity at 178 MHz and the core power at 5 GHz.}
\end{figure*}

\begin{table}
  \begin{center}
  \caption{Radio properties of HIGs and BLOs}
  \label{distmom}
  \begin{tabular}{c c|c c c }
    \hline \hline
Class & Parameter & Median & Average & sigma \\
    \hline
HIG &  log $L_{178}$  & 34.16$\pm$0.12 &  33.95$\pm$0.10 & 0.57 \\
BLO &  log $L_{178}$  & 33.98$\pm$0.14 &  34.04$\pm$0.12 & 0.49 \\
\hline
\multicolumn{2}{c}{      $\Delta$log$L_{178}$}& 0.18$\pm$0.19 &
-0.09$\pm$0.15 \\
    \hline
HIG &  log $R$       & -3.20$\pm$0.11 & -3.22$\pm$0.09 &  0.50$\pm$0.07 \\
BLO &  log $R$       & -1.97$\pm$0.24 & -2.13$\pm$0.19 &  0.81$\pm$0.14 \\
\hline
\multicolumn{2}{c}{      $\Delta$log$R$}& 1.23$\pm$0.27 & 1.09$\pm$0.21 \\
\hline
\end{tabular}\\
  \end{center}
  The luminosity at 178 MHz are in \ergsHz\ units; $R$ is the core dominance ($L_{core}/L_{178 \, \rm MHz}$). 
\end{table}

Several tests for the validity of the UM are based on the radio properties of
the objects considered. The basic requirement is that the distribution of
extended/low-frequency radio power of the two AGN classes considered do not
differ.  This is met by the 3CR sample. Indeed, from the point of view of
their radio power, the average values of $L_{178 \, \rm MHz}$ of the BLOs and
HIGs classes differ by only 0.18 dex and their median by -0.09 dex (see
Table \ref{distmom} and Fig. \ref{isto}) and also the spread of these
distributions are very similar (0.57 dex for HIG, 0.49 dex for BLO). We
verified through a Kolmogorov-Smirnov (KS) test that the two populations are
not different at a statistical significance level greater than 90\%.

According to the UM, the core dominance (defined as the ratio between the core
power at 5 GHz and the total luminosity at 178 MHz, i.e., $R = P_{\rm
  core}/L_{178 \, \rm MHz}$) should be larger in BLOs than in HIGs, because BLOs are
seen at smaller angles with respect to the jet direction than HIGs. This
produces a stronger relativistic Doppler boosting of the jet emission in BLOs,
causing their radio cores to be relatively brighter. Since the extended radio
emission is isotropic, only $P_{\rm core}$ could suffer from beaming
effects. Thus the core dominance $R = P_{\rm core}/L_{178 \, \rm MHz}$ is a good
estimator of beaming and orientation. This effect also provides us with a tool
to explore the jet properties in these radio sources.

The intensity of the core emission is enhanced, with respect to its intrinsic
value, $I_0$, as $I=I_0 \delta^{p+\alpha}$ where $\delta$ is the relativistic
Doppler factor. $\delta$ depends on the velocity of the jet ($v=\beta c$) and
on the angle $\theta$ formed with our line of sight as $\delta=\Gamma^{-1}
 (1-\beta {\rm cos} \theta)^{-1}$, where $\Gamma={(1-\beta^2)^{-1/2}}$ is
the bulk Lorentz factor of the jet, $\alpha$ is the spectral index in the band
considered ($\alpha \sim 0$ for the radio core); $p$ instead depends on the
structure of the emitting region ($p=2$ for a cylindrical jet and $p=3$ for a
single emitting blob, see Urry \& Padovani 1995).

\begin{figure*}
  \centerline{ 
    \psfig{figure=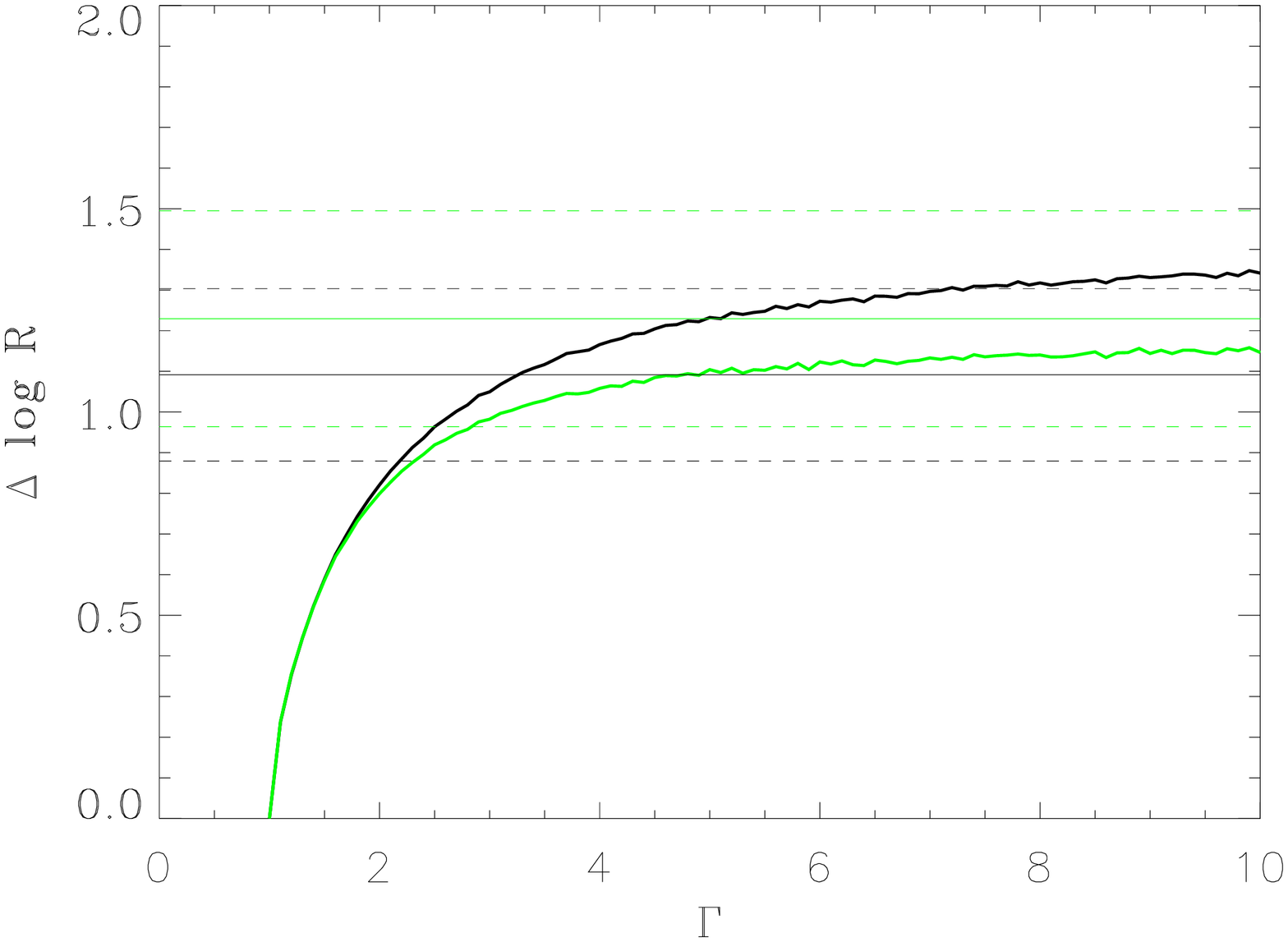,angle=90,width=0.55\linewidth}
    \psfig{figure=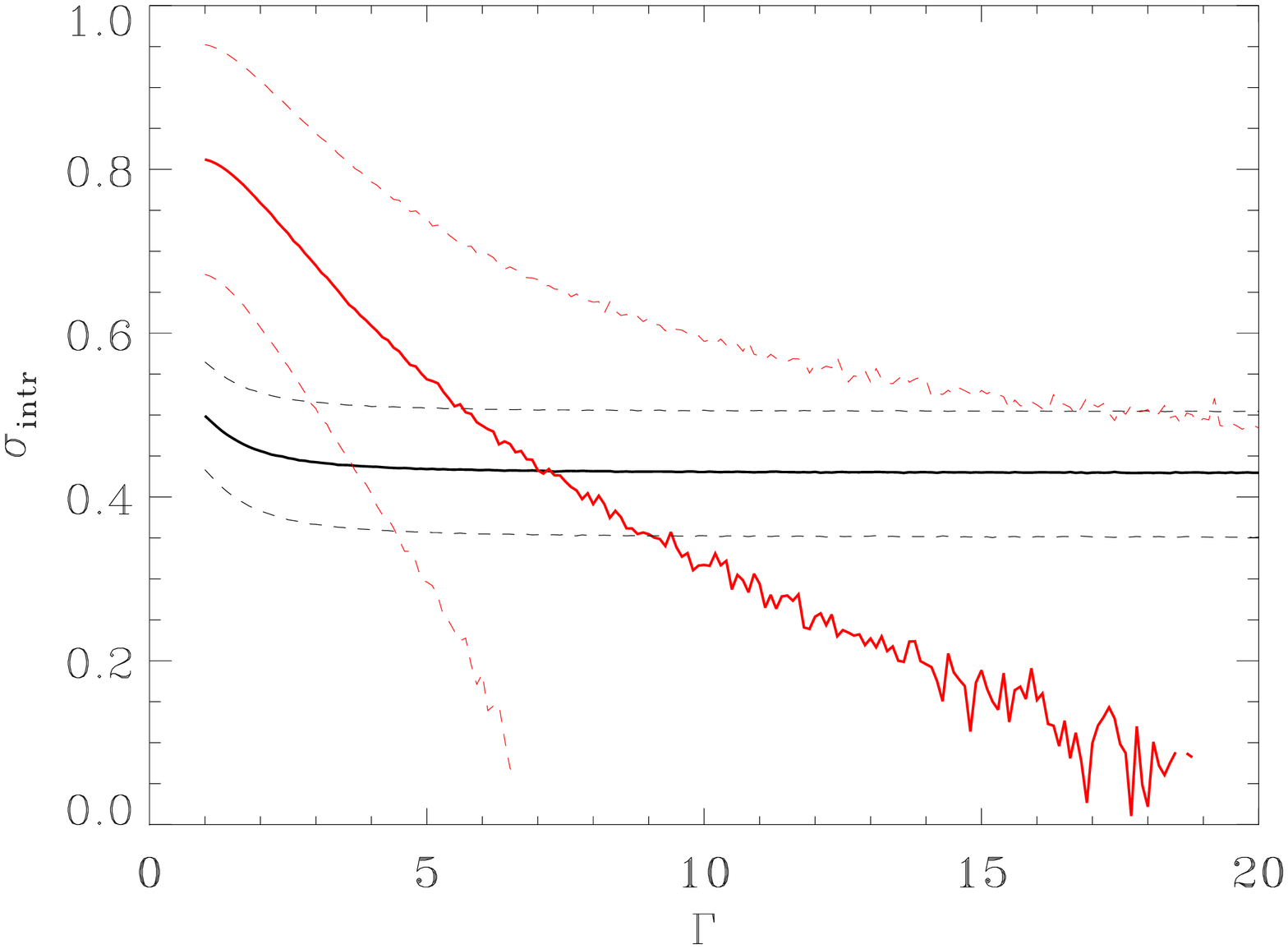,angle=90,width=0.55\linewidth} }
  \caption{\label{gamma} Results of Monte Carlo simulations to derive the jet
    Lorentz factor $\Gamma$ and the intrinsic spread of core dominance
    $\sigma_{\rm intr}$. The left panel shows the core dominance difference,
    $\Delta \rm{log} R$, of a simulated sample of HIGs and BLOs as a function
    of $\Gamma$. The black solid curve is for the average value, while the
    green line is for the median. The horizontal lines show the observed
    values, with the dashed lines representing the errors. Right panel: curves
    of the intrinsic spread of core dominance, $\sigma_{\rm intr}$, required
    to reproduced the observed values of the widths of the core dominance
    distributions for HIGs (black) and BLOs (red) as a function of
    $\Gamma$. At the intercept the constraints for HIGs and BLOs are
    simultaneously satisfied. The dashed lines for each class are obtained by
    considering the errors on the dispersions.}
\end{figure*}

We left out from the analysis three objects (namely, 3C~93.1, 3C~180, and
3C~458) because they do not have a 5 GHz radio core measurement in the
literature, leaving us with 48 objects.

The presence of beaming effects on the radio core is clearly shown by the
histograms of core dominance, see Fig. \ref{isto} (left panel). The median of
log $R$ is $-3.20$ and $-1.97$ for HIGs and BLOs, respectively (the average
values are instead $-3.22$ and $-2.13$, respectively) and thus they differ by
more than a factor of 10 (Table~\ref{distmom}). A KS test indicates that the
two populations are different at a level of confidence of $>$99\%.

The $R$ distributions for HIGs and BLOs are directly associated with the jet
properties. For example, the higher is $\Gamma$, the largest is the difference
in core dominance between the two classes. In the following we use the
observational information on the core dominance to constrain the jet
properties of FR~II radio galaxies with high-ionization optical spectra.

We start running a Monte Carlo simulation to study the relation between
$\Gamma$ and the separations between the core dominance distributions of HIGs
and BLOs. Operatively, we extract 100,000 objects oriented at a random angle
in the plane of the sky considering the case of a cylindrical emitting region
($p=2$). Using a torus angle of 50$^\circ$, as estimated above, we derive the
moments of the resulting core dominance distributions of HIGs and BLOs at
varying the jet Lorentz factor. In the left panel of Fig.~\ref{gamma}, we
consider the differences in the average $R$ between BLOs and HIGs. The
observed value of $\Delta $log$R = 1.09 \pm 0.21$ is reproduced for
$\Gamma=3.3^{+4.2}_{-1.1}$ . By using the observed difference of the median,
we obtain $\Gamma\gtrsim2.8$. By considering the possibility of a clumpy
torus, these results change only marginally: for $t=10^\circ$ we obtained a
slightly higher value, i.e., $\Gamma \sim 4.0$.

We then modeled the spreads of the core dominance distributions,
$\sigma_R$. These depend not only on $\Gamma$ but also on the presence of an
intrinsic difference in the core dominance among the various sources. We
assumed that the intrinsic distribution is described by a logarithmic gaussian
of width $\sigma_{\rm intr}$ and ran a Monte Carlo simulation. For each value
of $\Gamma$ we derived the corresponding value of $\sigma_{\rm intr}$ that
reproduces the observed values of $\sigma_R$ (0.50 for HIGs and 0.81 for
BLOs). With this procedure we obtain the curves (black for HIGs, red for BLOs)
shown in Fig. \ref{gamma} (right panel). At the location of the intercept
($\Gamma = 7.2$ and $\sigma_{\rm intr}=0.43$) the constraints for HIGs and
BLOs are simultaneously satisfied. Considering the errors on $\sigma_R$, we
found $4.4<\Gamma<18$ and $0.35< \sigma_{\rm intr} <0.51$.

These simulations, although instructive, do not completely exploit the
available observational data, i.e., the full distributions of core dominance
for the two classes. We then proceeded to a further simulation, extracting
randomly oriented samples of 48 objects (30 HIGs and 18 BLOs) and estimating
the differences between each pair of the ordered lists of observed and
simulated core dominance. We varied three parameters: the $\Gamma$ factor, the
intrinsic spread of core dominance, $\sigma_{\rm intr}$, and, in addition to
the previous discussion, also the intrinsic value of the core dominance,
$R_{\rm intr}$.  For each set of free parameters, the procedure is repeated
10,000 times, deriving the average value of the sum of the offsets
squared. The best fit corresponds to the set of parameters that returns the
minimum average value and it is obtained for $\Gamma=3.81^{+1.68}_{-1.11}$,
$\sigma_{\rm intr}=0.42^{+0.16}_{-0.13}$, $R_{\rm intr}=-2.38^{+0.43}_{-0.27}$. The uncertainties have been estimated
considering the range of the parameters for which the $\chi^2$ value increases
by 3.5.\footnote{This value corresponds to the amount that the $\chi^2$ is
  allowed to increase for a confidence level of 68\% and for three free
  parameters.}

We repeated the analysis adopting $p=3$ and obtained
$\Gamma=1.70^{+0.45}_{-0.21}$, $\sigma_{\rm intr}=0.52^{+0.11}_{-0.19}$,
$R_{\rm intr}=-2.89^{+0.13}_{-0.09}$.

\section{Size of the radio sources and the UM}
\label{step7}
\begin{figure}
  \centerline{ \psfig{figure=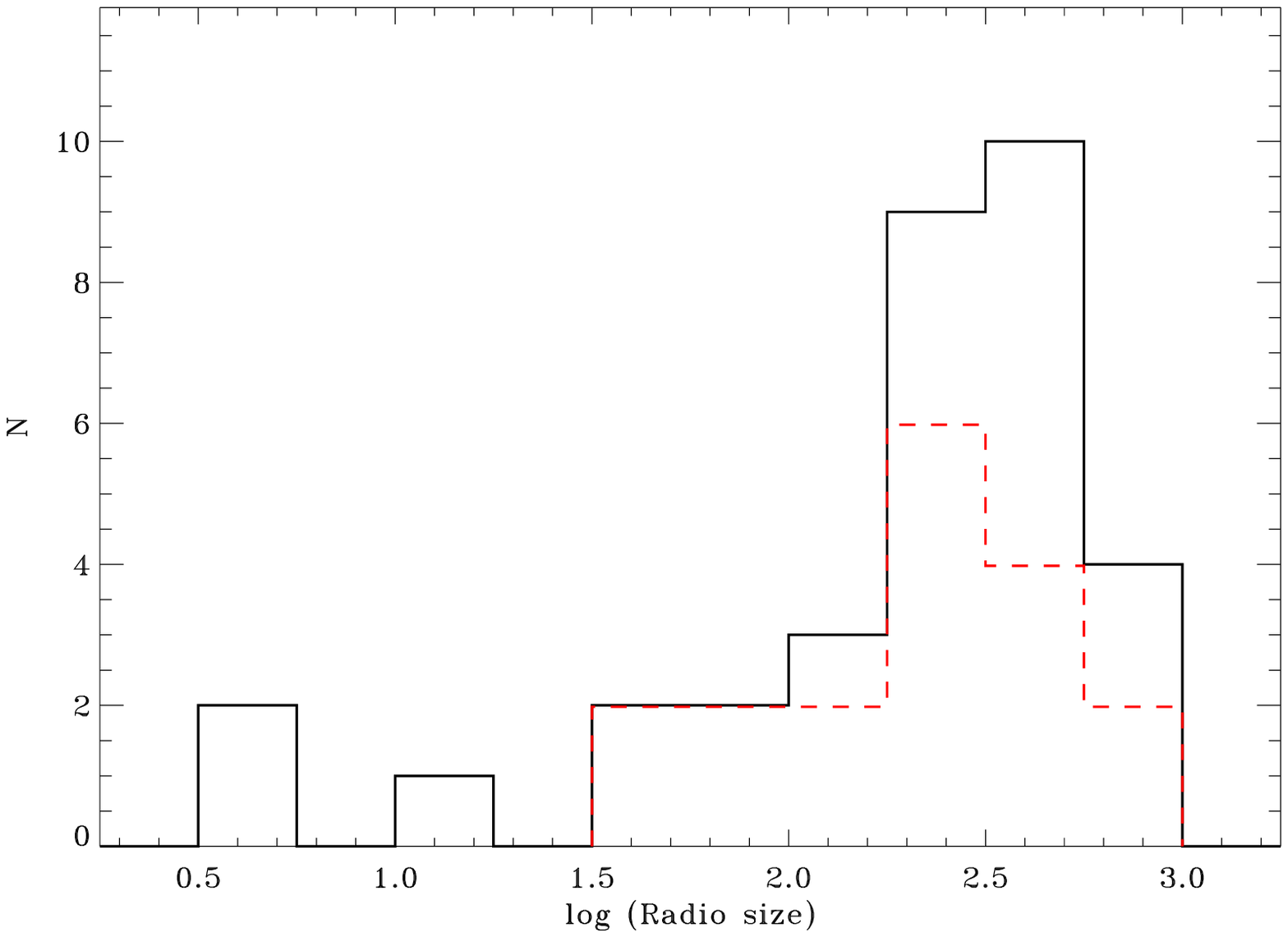,width=1.0\linewidth}
 }
 \caption{\label{size} Size distribution (in kpc) of the radio sources
   associated with HIGs (black histogram) and BLOs (red histogram)}
\end{figure}

In addition to the information on the total and core luminosities, the radio
emission provides us with a further test on the UM. If the HIGs and the BLOs
are two sub-samples of intrinsically identical sources, differing only for
their orientation, this should affect the size distribution of the two
classes. HIGs, being observed closer to the plane of the sky, should appear
larger than BLOs. The distributions of sizes of the two classes, derived from
literature images, are shown in Fig. \ref{size}. Both are very broad, with
most objects having sizes in the range 30 and 600 kpc. A KS test indicates
that the linear sizes of the two populations are not statistically different
at a $>$90\% level.

Having derived the critical viewing angle that separates HIGs from BLOs,
$\theta_{c}\sim50^\circ$, we can estimate the expected ratio between the sizes
of the two populations. The ratio between the median sizes of HIGs and BLOs is
expected to be 1.7. The observed median values are $\cal{S_{\rm HIG}}$ = 288
kpc and $\cal{S_{\rm BLO}}$ = 235 kpc for HIGs and BLOs, respectively.
However, the uncertainties on these values are rather large, 0.13 and 0.17
dex for the two groups, respectively, corresponding to a poorly constrained
ratio of the medians of $1.2^{+0.7}_{-0.4}$.

Thus, the test based on the radio sources on the UM is not conclusive since
the size distribution is dominated by the intrinsic scatter rather than by the
projection effects. 

\section{Narrow lines properties}
\label{step6}

\begin{figure*}
  \centerline{\psfig{figure=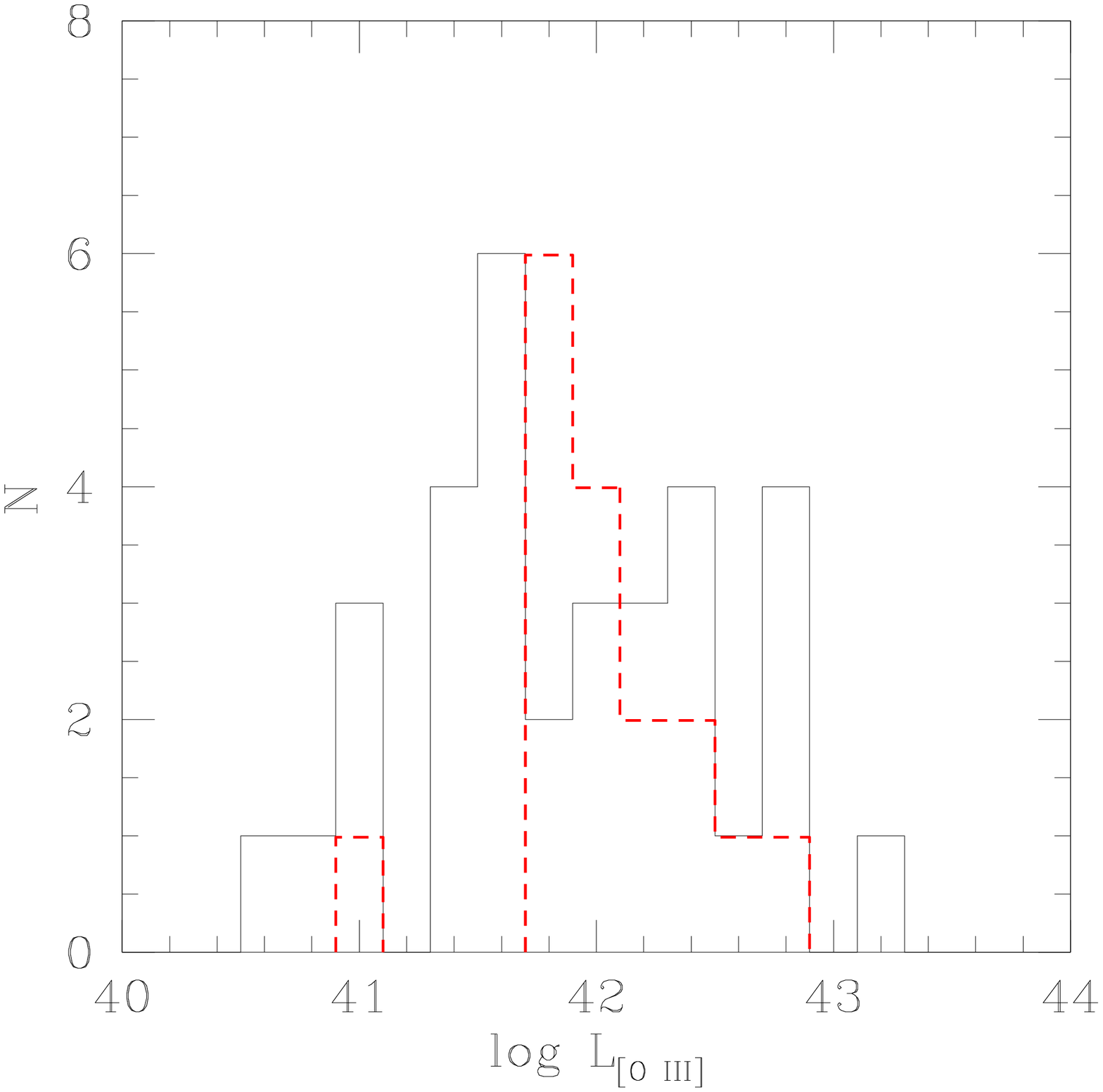,width=0.33\linewidth}
   \psfig{figure=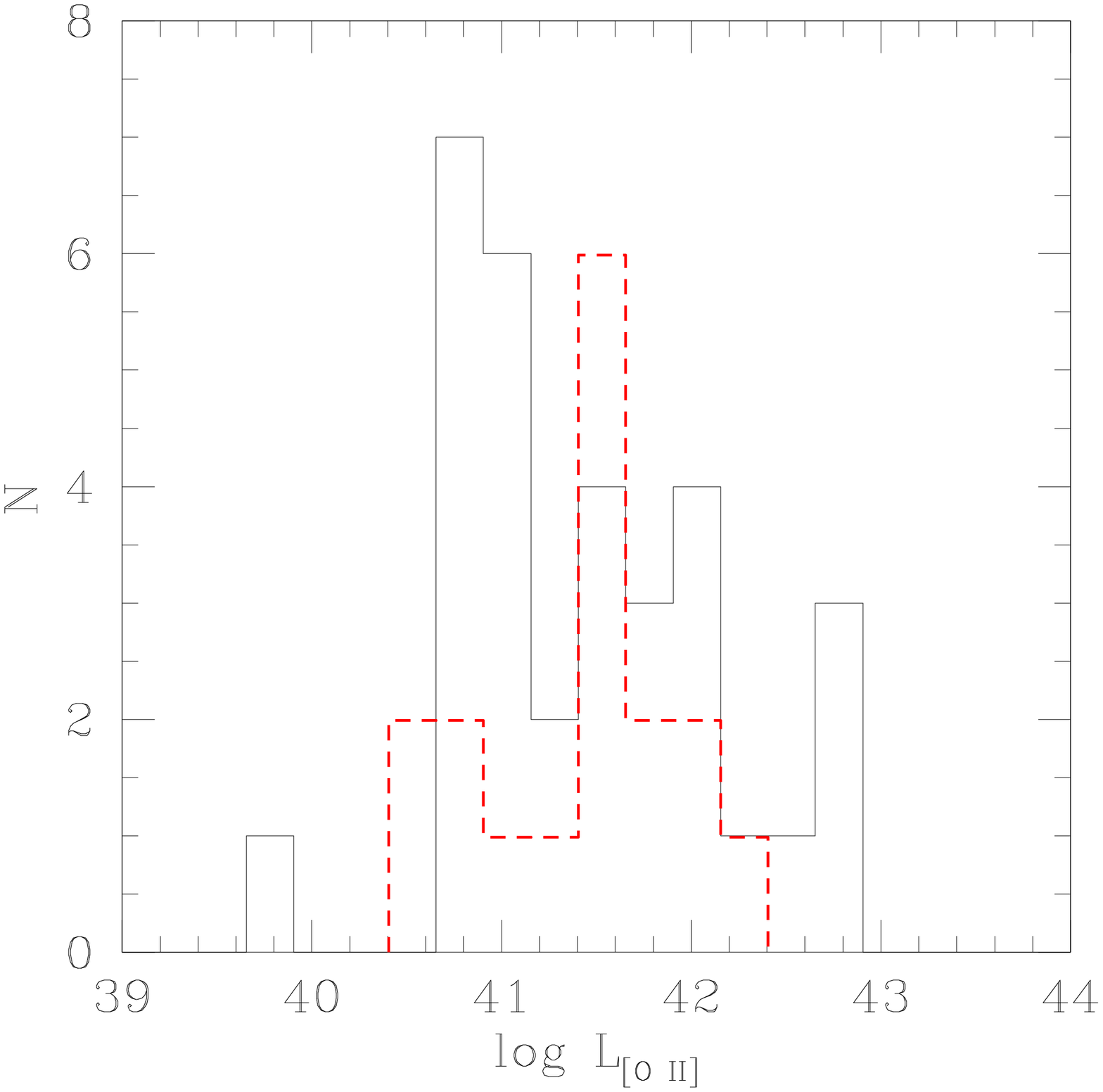,width=0.33\linewidth}
   \psfig{figure=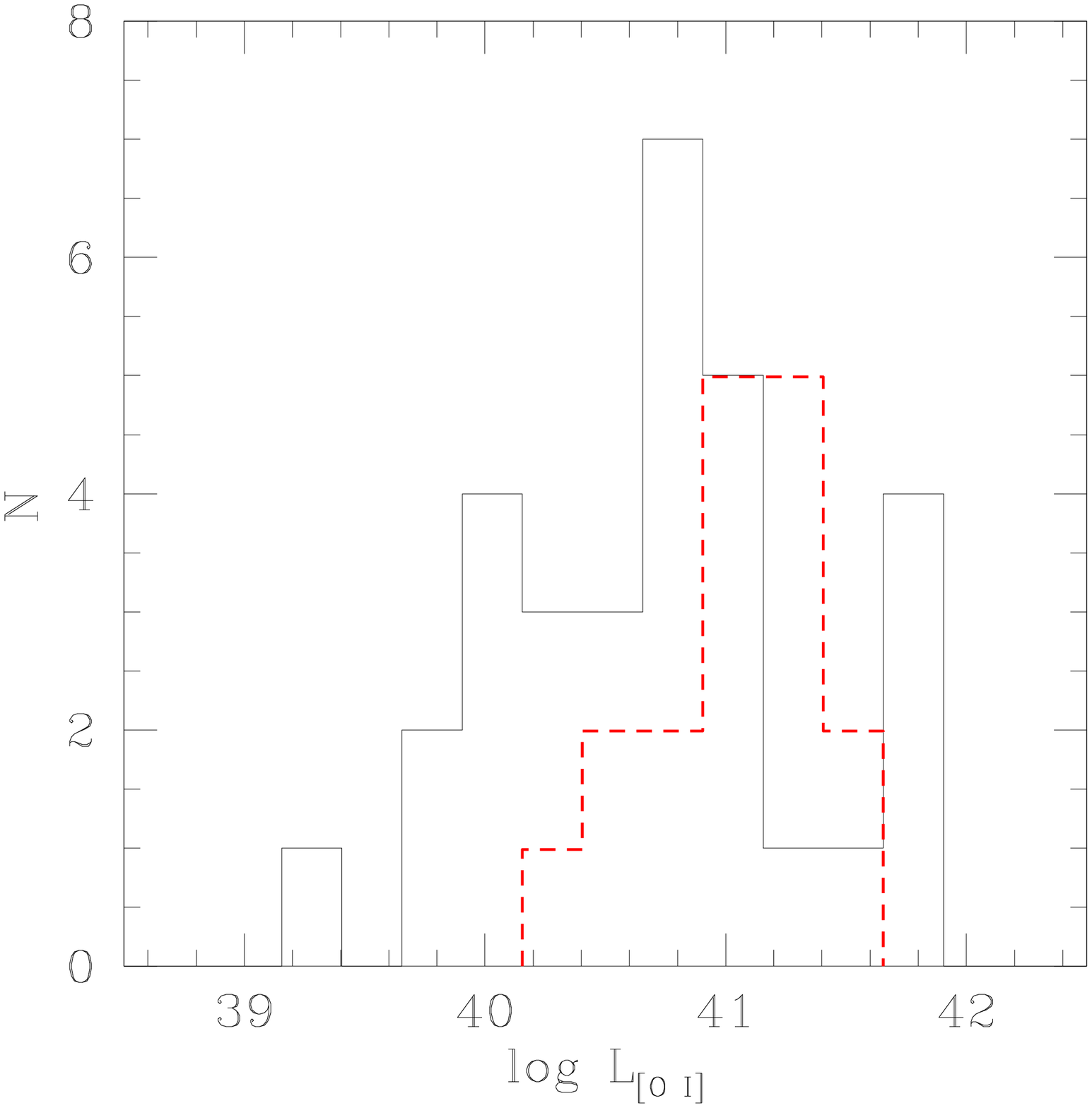,width=0.33\linewidth}
 }
 \caption{\label{istorighe} Histograms comparing the luminosity of HIGs (solid
   black) and BLOs (dashed red) in three narrow emission lines, from left to
   right, [O~III], [O~II], and [O~I], all in erg s$^{-1}$ units.}
\end{figure*}

In \citet{buttiglione10} we found that HIGs and BLOs lie in the same regions
of the optical spectroscopic diagnostic diagrams, indicating that they share
similar values of the main emission lines ratios. However, looking at the
diagrams in detail, we noticed that BLOs have relatively stronger [O~I] line,
by an average factor of $\sim 2$, than those of HIGs. We then explore this
issue in more depth by comparing the luminosity distributions of 3 brightest
oxygen optical emission lines, namely [O~III], [O~II], and [O~I], in HIGs and
BLOs (see Fig. \ref{istorighe}).

The median of the [O~III] and [O~I] distributions of the two classes differ
by a factor $\sim$ 2 (see Table \ref{distmomrighe}). A KS test indicates that
the two populations differ in $L_{\rm [O~III]}$ and $L_{\rm [O~I]}$ at a
statistical significance level greater than 95\%. Conversely, looking at the
[O~II] line, the moments of the distributions differ only marginally and the
differences in their cumulative distributions are not statistically
significative.

A similar result was already found by \citet{lawrence91} who noted that
  the [O~III] luminosity of narrow-lined objects was lower than in broad-lined
  objects at the same radio power. They observe this effect for various AGN
  samples selected in optical, infrared, X-ray, and radio band.  Furthermore,
  \citet{hes93} found for 3C radio galaxies that the [O~II] luminosities do
  not differ for narrow and broad-lined objects, again in agreement with our
  results.

These findings can be still accommodated within the UM by assuming that the
NLR is {\sl partially} obscured by the torus. The result that, beside the
known difference in [O~III], as well as the [O~I] luminosities, differ between
HIGs and BLOs, suggests that this might be due to a NLR density stratification
(rather than to an ionization stratification). Indeed, the three lines
considered are associated with different critical densities, with the [O~II]
lines having the lowest value.\footnote{The logarithms of critical densities,
  in cm$^{-3}$ units, are $\sim$ 3.5, 2.8, 5.8, and 6.3 for
  [O~II]$\lambda3726$, [O~II]$\lambda3729$, [O~III]$\lambda5007$, and
  [O~I]$\lambda6300$, respectively \citep{appenzeller88}.} It can be envisaged
that approximately half of the [O~III] and [O~I] emission is produced in a
compact, high density region (with a density exceeding the [O~II] critical
density, i.e., $\gtrsim 10^3$ cm$^{-3}$) located within the walls of the
obscuring torus.

\begin{figure}
  \centerline{ \psfig{figure=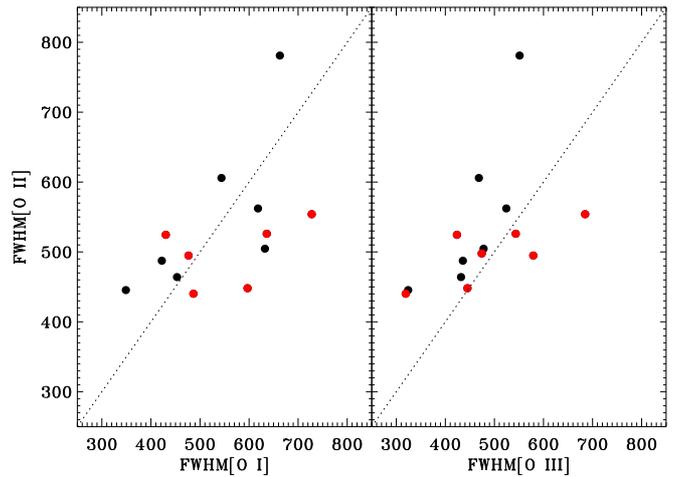,width=1.0\linewidth}
 }
 \caption{\label{o1o2o3} Comparison of the widths (in km s$^{-1}$) of three
   oxygen emission lines measured from the available 14 SDSS spectra. BLOs are
   the red circles, HIGs are represented by the black squares.}
\end{figure}

In line with this scenario, other NLR properties should differ between BLOs
and HIGs. In particular the widths of narrow lines of higher critical
densities should be broader in BLOs since they have a larger contribution from
emission originating closer to the central black hole. This can be tested by
measuring the line widths of the three oxygen lines. The spectral resolution
of the TNG data obtained for this survey is not suited for such an
analysis. We then limit ourselves to the 14 sources (7 BLOs and 7 HIGs) with
available SDSS spectra. The velocity widths derived from a single gaussian fit
to the lines are shown in Fig. \ref{o1o2o3}. Although the number of sources is
rather limited, we find tentative evidence that indeed BLOs have broader [O~I]
and [O~III] lines with respect to HIGs of similar [O~II] width. This supports
the idea of a radial density gradient in the NLR.

\begin{table}
  \begin{center}
  \caption{HIGs and BLOs narrow lines distributions}
  \label{distmomrighe}
  \begin{tabular}{c c|c c c }
    \hline \hline
Class & Line & Median & Average   &                   \\
    \hline                                              
HIG &  [O~III] & 41.85 $\pm$ 0.14 &  41.89 $\pm$ 0.11 \\ 
BLO &  [O~III] & 42.03 $\pm$ 0.13 &  42.03 $\pm$ 0.10 \\ 
    \hline                                              
HIG &  [O~II]  & 41.43 $\pm$ 0.15 &  41.45 $\pm$ 0.12 \\ 
BLO &  [O~II]  & 41.50 $\pm$ 0.15 &  41.42 $\pm$ 0.12 \\ 
    \hline                                              
HIG &  [O~I]   & 40.77 $\pm$ 0.29 &  40.49 $\pm$ 0.23 \\ 
BLO &  [O~I]   & 41.07 $\pm$ 0.11 &  41.01 $\pm$ 0.09 \\ 
    \hline
\end{tabular}\\
  \end{center}
Logarithms of the line luminosities in $\ergs$ units.
\end{table}

\section{Low ionization galaxies and the unified model}
\label{lstep}

\begin{table}
  \begin{center}
  \caption{The sample of the 3CR FR~II/LEGs with z$<$0.3.}
  \label{leg}
  \begin{tabular}{l|c | c c }
    \hline \hline
 Name &  z &  Log P$_{core}$ &  Log $L_{\rm 178}$  \\
    \hline
3C~015   &  0.0730   & 31.64    &  33.30   \\
3C~88     &0.0302     &  30.57   &   32.49    \\
3C~123   &  0.2177   &   32.00  &    35.41    \\
3C~132   &  0.2140   &   31.58  &    34.25     \\
3C~153   &  0.2769   &   29.94  &    34.56     \\
3C~165   &  0.2957   &   31.30  &    34.57     \\
3C~166   &  0.2449   &   32.92  &    34.42     \\
3C~173.1&  0.2921   &   31.39  &    34.61     \\
3C~196.1&  0.1980   &   30.72  &    34.31     \\
3C~213.1&  0.1940   &   31.15  &    33.84     \\
3C~236   &  0.1005   &   31.62  &    33.56     \\
3C~288   &  0.2460   &   31.73  &    34.53     \\
3C~310  &  0.05350  &    30.72 &     33.56     \\
3C~326  &  0.08950  &    30.45 &     33.60     \\
3C~349  &   0.2050   &   31.35  &    34.20     \\
3C~353  &  0.03043  &    30.61 &     33.69     \\
3C~357  &   0.1662   &   30.63  &    33.86     \\
3C~388  &  0.09100  &    31.15   &   33.70     \\
3C~401  &   0.20104 &     31.67  &    34.38     \\
3C~430 &   0.05410  &    30.06   &   33.36     \\
3C~460  &   0.2690   &   31.59   &   34.26     \\
   \hline
 \end{tabular}\\
 \end{center}
Radio luminosities in erg s$^{-1}$ Hz$^{-1}$ units.
\end{table}

Until now, we only considered radio sources characterized by emission lines of
high ionization, that represent less than half of the 3CR sample up to
$z=0.3$. However, the study of HIGs and BLOs provides us with a useful
benchmark to explore the properties of the other main spectroscopic class of
sources in the sample, i.e., the low ionization galaxies (LIGs). While many
LIGs have a FR~I radio morphology, our spectroscopic study
confirms the presence of a significant number of LIGs (21, see Table
\ref{leg}) with a clear FR~II structure, despite our rather strict criteria
for the definition of the FR type.

The radio properties of FR~II/LIGs are generally similar to those of HIGs and
BLOs. Beside the morphology, they cover the same range of radio power and also
have a similar distribution of core dominance \citep{buttiglione10}. The
questions that arise are: which is the link between low and high ionization
galaxies? Which role do LIGs play in the UM? How do the properties of the LIGs
jets compare with those of HIGs and BLOs? We already know from previous
studies (e.g., \citealt{laing94}) that the core dominance distribution of 3CR
FR~II LIGs indicates a randomly-oriented population, different from the
high-ionization galaxies. Therefore, the study of the complete 3CR sample,
precisely its LIG sub-sample, can return more solid results on such
questions.

\begin{figure*}
  \centerline{ 
    \psfig{figure=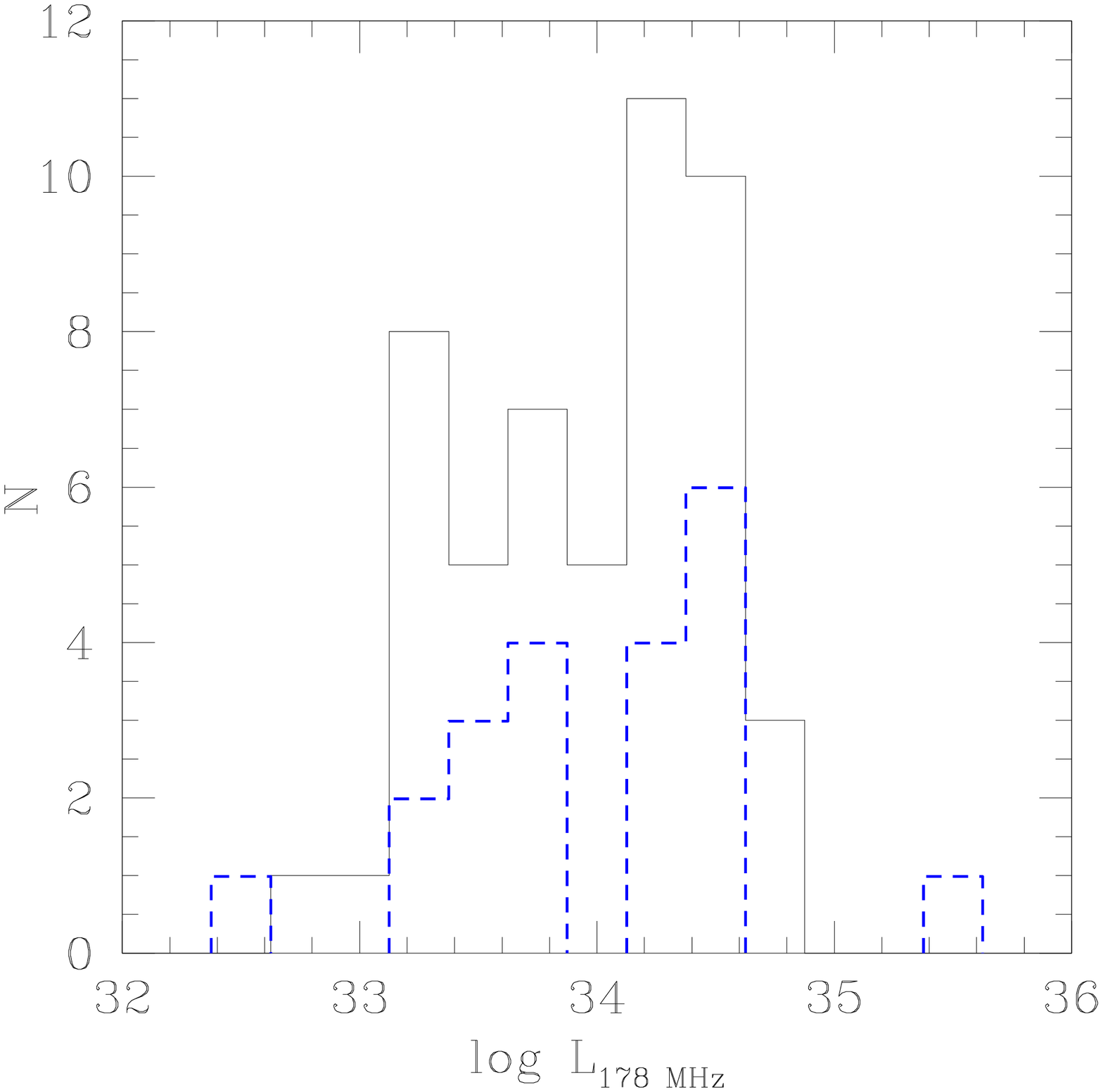,width=0.50\linewidth}
    \psfig{figure=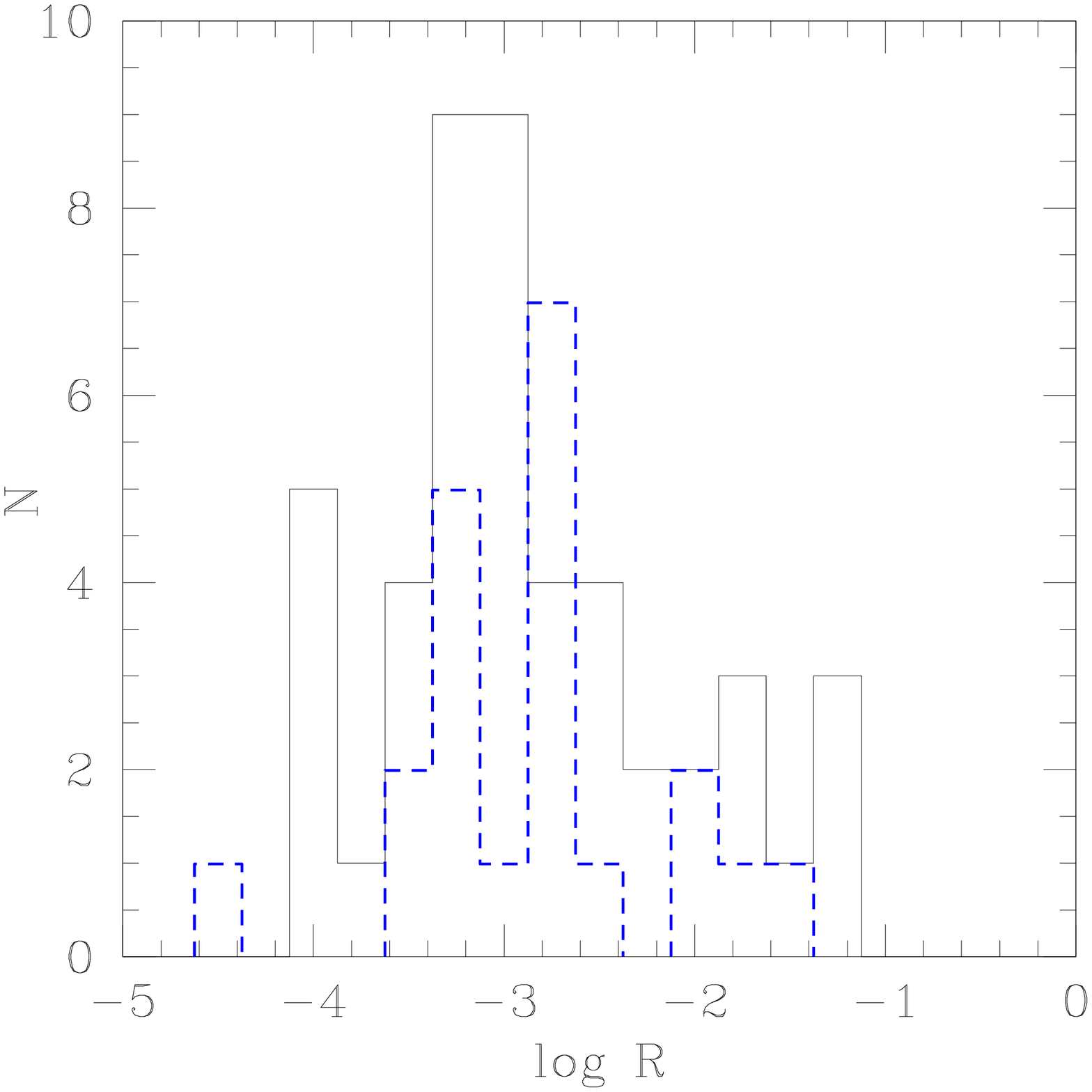,width=0.50\linewidth}}
  \caption{\label{istoleg} Comparison of the distributions of total radio
    luminosity at 178 MHz (left) and of the core dominance $R$ (right) for
    FR~II/LIGs (dashed blue) and HIGs+BLOs (solid black).}
\end{figure*}

First of all, we tested with a KS test that the distributions of radio power
of LIGs and HIGs/BLOs are not statistical different (see Fig. \ref{istoleg},
left panel). This leaves open the possibility that LIGs might represent a
third group of radio sources, part of the same unification scheme with BLOs
and HIGs. Furthermore, in the light of the results of the previous section,
the differences in optical line ratios between LIGs and HIGs might be due to
selective obscuration. For example, LIGs might be seen along a line of sight
close to the equatorial plane of the torus, causing the obscuration of a
substantial fraction of their NLR.  This might account for the factor of $\sim
10$ deficit in their [O~III] line with respect to HIGs and BLOs
\citep{buttiglione10} and also for the differences in line ratios. The optical
spectra of FR~II/LIGs never show the presence of a significant BLR, and this
also would be naturally explained if they are all observed at large
$\theta$. 

In this framework, the UM provides a clear prediction of the radio properties
of LIGs: they must show a lower core dominance and a narrower distribution of
$R$, than those of HIGs. But this is not the case: the distribution of $R$ for
LIGs has a median value of log $R$ = -2.84 and a spread of 0.70 dex,
substantially larger than in HIGs (for which log $R$ = -3.20 and
$\sigma_{R}=0.57$). Thus LIGs cannot be objects intrinsically identical to
HIGs and BLOs just seen at larger viewing angles. In addition, the
distribution of core dominance of LIGs is instead not statistically
distinguishable from the distribution of the population formed by HIGs plus
BLOs (see also Fig. \ref{istoleg}, right panel) at a 90\% level.

Since broad lines are intrinsically absent in LIG spectra, no indications on
orientation can be obtained from their optical spectra. To explore their
properties, we must rely only on their radio properties. Following the
analysis described in Sect. \ref{step5}, we modeled the distribution of $R$ in
LIGs and we derived the following set of parameters: $\Gamma < 3.25$,
$\sigma_{\rm intr}<0.74 $, $R_{\rm intr}=-2.90^{+0.34}_{-0.21}$ for
$p=2$.\footnote{For $p=3$ we find $\Gamma < 1.90$, $\sigma_{\rm intr}<0.70 $,
  $R_{\rm intr}=-3.02^{+0.27}_{-0.13}$.} Due to the small number of objects,
the constraints on the jet's parameters in LIGs are rather weak.

 \section{Discussion}
\label{discap5}

The complete 3CR optical survey allows us to derive more accurate results
  than in previous works in the framework of testing the predictions of the UM
  for FR~II radio sources, when high and low ionization objects are properly
  separated. In particular, the 3CR HIGs and BLOs, limited at $z<0.3$, show
  differences and similarities in the radio and line-emission properties which
  are ascribable to a random orientation of the parent population, fundamental
  requirement of the UM.

A key element in the UM is the presence of a circumnuclear structure that
hides the view of the innermost regions of the AGN in the optical band when
seen at large angle from its axis. Our spectroscopic survey clearly cannot
provide any direct evidence for the existence of an obscuring torus. However,
the connection between the radio core dominance and the optical spectra (i.e.,
to the presence/absence of broad emission lines), confirms the link between
orientation of the AGN and its spectral classification, requiring the presence
of an absorbing structure.

In the assumption that the torus is present in FR~II-type radio-loud AGN,
  we can obtain information on its geometric structure. The simple torus
  geometry we assumed provides an estimate of its opening angle \tc =
  50$^{\circ} \pm 5$.  Although the sample is limited to $z=0.3$, there is not
  evidence of change of the torus angle throughout the redshift and luminosity
  range of the sample. Furthermore, \citet{barthel89} selected a sample of
  $\sim 50$ quasars and radio galaxies in the 3C catalog with $0.5 <z< 1.0$
  with a radio power in the range $L_{178 \, \rm MHz} \sim 10^{35}-10^{36}$
  \ergsHz. He derived a value of \tc $= 44^{\circ}.4$, in agreement with our
  estimate. This further supports our conclusion that the torus geometry does
  not vary significantly with radio power over $\sim$3 orders of
  magnitudes. This contrasts with the 'receding torus' model proposed by
  \citet{lawrence91} and \citet{hill96} who find a decreasing fraction of
  broad and narrow line objects in the 3CR at increasing radio
  luminosity. Nevertheless, the problem is far to be solved since there are
  conflicting results in literature. In fact, another study on 3CR radio
  sources which compares objects at $z<0.3$ with others to $z\sim 1$ returns
  that the covering factor of the obscuring structure decreases as the
  redshift increases \citep{varano04}. The solution might be reached with
  large samples which cover wide range of luminosities and redshifts. In
  addition, the inclusion of LIGs in the composite sample might strongly
  affect the results, since they appear to be a different FR~II population
  from HIGs and BLOs.

We also consider the possibility of a clumpy structure for the torus. This
allows us to see directly the nuclear regions according to an arbitrary
probabilistic law, that is not null even along the torus equator. Limiting to
a level of torus porosity allowed by the infrared observations of AGN, we find
that the torus opening angle might decrease only slightly, to $\sim
43^{\circ}$, and this has a marginal effect on the estimates of the jet
Lorentz factor. However, the question on the nature of the torus is still far
from being resolved as suggested by the conflicting results on radiative
transfer in clumpy media (e.g., \citealt{landt10}).

The test of the UM based on the radio source size of these 3CR source is
instead not conclusive. This is due to the fact that the expected
foreshortening of BLOs with respect to HIGs amounts to only a factor of 1.7
and is too small to be appreciated in the observed size distribution dominated
by the intrinsic scatter. This effect should be derived by comparing two {\it
  intrinsic} size distributions that are very broad, ranging from $\sim$30 to
$\sim$600 kpc, leading to an error of the median sizes of the two classes of a
factor of $\sim$ 3. Only a substantially larger number of objects might unveil
a genuine difference in size between HIGs and BLOs. The absence of this
foreshortening of the sizes of quasars as compared to those of radio galaxies
of similar flux densities or at similar (low and high) redshifts has been
observed in larger samples (e.g. \citealt{boroson11,singal13,boroson13}). A
possible reason of this conflict with what expected from the UM can be
ascribed to the presence of LIG in those large sample which do not participate
in orientation unification schemes as explicitly discussed by
\citet{singal13}. \citet{dipompeo13} underlines the importance of considering
the intrinsic size distribution of radio sources in this context. They found
that, while it is possible to reconcile conflicting results purely within a
simple, orientation-based framework, it is very unlikely.

Adopting the UM model, it is possible to constrain the jet properties needed
to reproduce the observed distributions of radio core dominance in HIGs and
BLOs. We find that this is obtained for a jet bulk Lorentz factor in the range
$\Gamma \sim 3 - 5$ (for $p=2$), nearly independent on the torus properties. A
comparison with previous results on the jet bulk Lorentz factor for radio
galaxies is necessary.  \citet{padovani91} derived an estimate of $\Gamma_{o}$
in the optical band comparing the number density of BL Lac objects and the
parent population of FR~I galaxies, by combining a correction for the effect
of beaming and contamination from the host galaxy on the optical luminosity
function. They found $8< \Gamma_o < 20$, a larger range than the Lorentz
factor inferred for the X-ray emitting plasma $\Gamma_X \sim 3$
\citep{padovani90} and roughly of the same order of the radio Lorentz factor,
$5 \lesssim\Gamma_r \lesssim 35$ \citep{urry91}. Studies on the Chandra X-ray
jets for FR~II radio galaxies provides measurements of the bulk Lorentz factor
in the range $2-7$ \citep{hardcastle02,siemiginowska03,hogan11}.  Whereas the
previous estimates of the bulk Lorentz factors are derived from luminosity
functions or emission models, \citet{hardcastle99b} takes in account the
information of the angle orientation from the optical spectroscopic
classification (into HIGs and BLOs) for a sample of FR~IIs, excluding LIGs. By
modeling the distributions of jet prominence of the sample, they derive a bulk
Lorentz factor of $\sim 5$. Conversely, there are no proper measurements of
$\Gamma$ for a sample of FR~II LIGs in the literature. 

The Lorentz factors inferred from the apparent superluminal jet motions
  (up to $\gamma\sim30$, see review from \citealt{kellermann07}), are
  substantially larger than our values. This issue was already argument of
  debate since the dynamic range of core dominances expected from superluminal
  motions predicts a much larger number of debeamed sources than what obtained
  from observations. This inconsistency can be ascribed to the simplistic
  assumption of a cylindrical jet model without any internal velocity
  structure.  This does not conform to some jet features, such as jet bending
  (e.g., \citealt{readhead83}). Instead, the inclusion of shock waves in the
  hydrodynamic models seems to better reconcile with the observations
  \citep{blandford84,lind85}. This result is reminiscent of what is found from
  the comparison of the properties of FR~I radiogalaxies and BL~Lacs
  \citep{chiaberge00}. Summarizing, the derivations of the bulk velocity of
  the relativistic jets in different bands and with approaches appear to be
  approximatively consistent with each other, even though the sources show
  different radio morphologies (FR~I and FR~II) and nuclear properties (HIGs
  and LIGs).

  Another key test of the UM is based on the properties of NLRs: They are
  structures extending out to a scale of several kpc and thence they are
  thought generally to be unaffected by nuclear obscuration. For this reason
  the luminosity of the narrow emission lines apparently represents an
  isotropic quantity, particularly useful when testing the UM. Instead, thanks
  to the completeness of the spectroscopic data we found evident discrepancies
  in terms of properties of the three brightest oxygen optical emission line
  ([O~III], [O~II], and [O~I]) between HIGs and BLOs.  Such differences are
  not in agreement with the UM at the zeroth-order approximation, which
  predicts similar NLR properties. However, our results can be accommodated
  within the UM, if we assume a partial obscuration by the torus on the
  nuclear region of the NLR. Furthermore, the luminosity distribution and the
  FWHM of the oxygen lines return an interesting result: they invoke a
    NLR density stratification, where an innermost region, compact and dense,
    is responsible to produce approximately half of the [O~I] and [O~III]
    emission lines. Their higher critical densities with respect to the [O~II]
    might account for a higher spatial concentration closer to the black
    hole.  This region appears to be a crucial component of the NLR,
  apparently confined to a scale of the order of the dust sublimation radius
  (i.e., pc scale, \citealt{barvainis87}), larger but comparable to the BLR
  radius \citep{bentz13}. It is tempting to associate it with the walls of the
  torus, indeed rich of dense neutral gas, producing the observed large amount
  of high velocity [O~I] emission. Its density is sufficiently low to allow
  the production of narrow lines, and thence lower than in the BLR, but some
  of them are already strongly depressed by collisional de-excitation. 
    Furthermore, in agreement with our partially obscured NLR model,
    spectro-polarimetric data show evidences of a prominent contribution from
    scattered [O~III] lines, emitted behind this obscuring material
    \citep{diserego97,cohen99}.

This applies not only to radio-loud AGN, but also to radio-quiet objects. In
fact \citet{lawrence87} show a similar line mis-match between type 1 and type
2 AGN. Direct evidence for a compact and dense NLR component comes from the
detection of variability of the \oiii\ flux in NGC~5548 over a timescale of a
few years \citep{peterson13}; they derived a size of a few pc and an
electronic density of $\sim 10^5$ cm$^{-3}$.  A similar structure is also seen
around the nuclei in FR~I radio galaxies from the analysis of HST spectra and
narrow-band images \citep{capetti05b}.

Finally, the study of the radio and core dominance distribution of 3CR FR~II
LIGs indicates that they do not have a preferred orientation in the plane of
the sky, supporting the results of \citet{laing94}. In this case, broad lines
are intrinsically missing in LIGs, at odds with what is seen in the BLOs/HIGs
class, and similarly to the indications obtained for the FR~I/LIGs (e.g.,
\citealt{chiaberge:ccc,baldi10b}).

The differences between low and high ionization radio galaxies cannot be just
due to orientation but they are intrinsically different objects at the nuclear
scales. Their different properties reflect the bimodality of accretion
  modes in RL AGN population.  While HIGs/BLOs show thermal nuclei
  characteristic of 'standard' accretion mode, LIGs appear to have
  synchrotron-dominated nuclei powered by radiatively inefficient accretion
  disk (e.g., \citealt{chiaberge02,hardcastle06,baldi10b}). The infrared
  emission is crucial to unveil the presence of a hidden quasar and separate
  the two AGN classes; this spectra band acts as a calorimeter in which a
  large fraction of the AGN bolometric power is reprocessed (see the review on
  this topic by \citealt{antonucci12}).

\section{Summary and conclusions}
\label{concl5}

The complete optical spectroscopic survey for 3CR sources with $z<0.3$
  \citep{buttiglione10} provides us with a homogenous sample perfectly suited
  to test various predictions and discuss the implications of the UM for FR~II
  radio galaxies. We exclusively consider the FR~II HIG sample which
consists of 33 narrow line objects and 18 broad-lined objects (BLOs). The aim
is to derive the main quantities involved in the UM for RL AGN: the opening
angle of the obscuring torus and the bulk Lorentz factor of the relativistic
jets. The method used is the study of the core dominance distribution of the
sample and their emission line properties. The main results are summarized as
follows:

\begin{itemize}

\item The HIGs/BLOs number ratio corresponds to an opening angle of an
  obscuring torus of \tc = $50^{\circ} \pm5^{\circ}$. There is no evidence of
  a dependence of this value with redshift/luminosity within the sample
  considered, up to $z=0.3$.

\item While HIGs and BLOs share the distribution of total radio luminosity,
  their core dominance distributions are significantly offset, by $\sim$ 1
  order of magnitude. This implies that it exists a strong link between the
  optical and radio properties, with the jets in BLOs forming an angle with
  the line of sight smaller than HIGs, supporting the validity of the UM. We
  modeled the distributions of $R$ to estimate the jet bulk Lorentz factor,
  obtaining $\Gamma \sim 3-5$.

\item We consider the possibility of a 'clumpy' torus: this has only a small
  impact on its 'opening angle' and on the value of $\Gamma$.

\item The test of the UM based on the radio source size is not conclusive, due
  to the small number of objects considered.

\item While the properties of the [O~II] emission line are similar in BLOs and
  HIGs, they differ for the \oiii\ and [O~I] lines. In particular, these lines
  are broader and more luminous in BLOs. This is consistent with a combination
  of obscuration and density stratification in the NLR. Approximately
    half of the line emitting gas (with high critical density, i.e. [O~I] and
    [O~III]) is located within the walls of the obscuring
    torus and it is visible only in BLOs, while it is obscured in HIGs.
 
\item Considering now the FR~II LIGs, they might be, in principle, all objects
  seen at high inclination, with the BLR and also most of the NLR hidden from
  our view. This is incompatible with their broad core dominance distribution
  that is instead consistent with what is expected from a sample of random
  oriented sources. Thus LIGs can not belong to the same UM with HIGs and
  BLOs. This result lends further support to the idea that LIGs constitute a
  separate class of radio-loud AGN, as already suggested by the differences in
  their nuclear properties. Unfortunately, due to the limited number of LIGs,
  we cannot perform a robust comparison between the jet Lorentz factor of the
  different classes of radio sources. Thence we are not able to conclude
  whether there is an association between the jet properties (and the jet
  launching mechanism) and the different spectral types.

\end{itemize}

Overall, the results obtained for FR~II radio sources are consistent with a
pure orientation-based unified model, when considering separately objects of
high and low ionization. Indeed, we find that the distributions of total radio
luminosity of HIGs and BLOs are not statistically distinguishable, while BLOs
have a higher radio core dominance than HIGs. This links the orientation
indicator based on the radio data with the effects of selective nuclear
obscuration, as expected in the UM framework. 

We find significant differences in the properties of narrow emission lines
between HIGs and BLOs, but this does not contrast with the UM even in its
simplest version. While historically the narrow lines have been considered
isotropic, there is mounting evidence that a significant fraction of the
emission from forbidden lines originates from a compact and dense region
located within the walls of the torus. This is visible only in type 1 AGN,
i.e. in the BLOs, and it is not exclusively associated to radio-loud objects,
being present also in radio-quiet AGN.

\begin{acknowledgements}

  R.D.B. acknowledges the financial support from SISSA, Trieste. We are
  grateful to the referee R. Antonucci for the extremely useful comments to
  improve the paper. This work is primarily based on the COSMOS data.

\end{acknowledgements}

\bibliography{my.bib}

\begin{appendix}
\section{Estimate of the orientation of a radio source based on the core dominance}

One interesting application of the derivation of the radio parameters (jet
bulk Lorentz factor $\Gamma$, intrinsic core dominance $R_{intr}$, and
intrinsic spread of the core dominance distribution $\sigma_{intr}$) from the
core dominance distribution of 3CR/FR~II radio galaxies (HIGs and BLOs) is the
possibility of estimating their orientation starting from the measurement of the
core dominance of individual objects. By inverting the link between the core
dominance $R$ and the viewing angle $\theta$ the analytical
relation is the following:

$$cos \theta = \frac{\Gamma -  \left(\frac{R_{intr}}{R}\right)
^{1/p}}{\sqrt{\Gamma^{2} - 1}} $$

\begin{figure}
  \centerline{ \psfig{figure=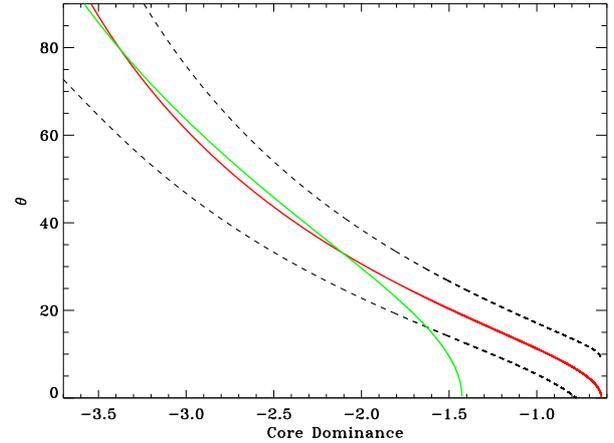,angle=90,width=0.99\linewidth}
 }
 \caption{\label{thetacd} The orientation angle vs. core dominance relation
   obtained by using the derived three parameters (jet bulk Lorentz factor
   $\Gamma$, intrinsic core dominance $R_{intr}$, and intrinsic spread of the
   core dominance distribution $\sigma_{intr}$) for HIG and BLO from their
   core dominance distribution. The red curve is for p=2 (cylindrical jet)
   with the error curves represented by the dashed line. The green curve is
   for p=3 (single emitting blob).}
\end{figure}

In Fig.~\ref{thetacd} we show the two curves obtained for $p=2$ and $p=3$ and
adopting the jet's parameters derived in Sect. \ref{step5}. We
also show the curves obtained by considering the errors on the three
parameters (see Section~\ref{step5}). For example, for a radio source with a
core dominance log$R = -2$ we derive $\theta=20^\circ\pm8^\circ$.

A possible application of such relations is the deprojection of a radio source
and the estimate of its real size.

\end{appendix}

\end{document}